\documentclass[journal]{IEEEtran}
\usepackage{cite}
\usepackage{mathtools}
\usepackage{amsthm,amsmath,amssymb}
\usepackage{amssymb}
\usepackage{mathrsfs}
\usepackage{subfigure}
\usepackage{graphicx}
\usepackage{epstopdf}
\usepackage{float} 
\usepackage{multirow}
\usepackage{booktabs}
\usepackage{url}
\usepackage{bm}
\usepackage{array} 

\usepackage[dvipsnames]{xcolor}
\usepackage{blkarray}
\graphicspath{{./figure/}}
\DeclareGraphicsExtensions{.pdf,.jpeg,.png,.jpg}

\usepackage{colortbl}
\usepackage[ruled,lined,linesnumbered]{algorithm2e}
\usepackage{caption}
\usepackage{subcaption}

\usepackage[hidelinks]{hyperref}
\usepackage{booktabs}
\usepackage{tikz}
\usepackage{tabularx}

\ifCLASSINFOpdf
\else
\fi

\graphicspath{{figures/}}
\usepackage{amsmath} 
\usepackage{ragged2e}

\usepackage{xcolor}
\usepackage{eso-pic} 
\definecolor{lightblue}{RGB}{230,245,255}

\newcommand{\TopNotice}{%
  \AddToShipoutPictureFG*{%
    \AtPageUpperLeft{%
      \raisebox{-1cm}[0pt][0pt]{
        \begin{minipage}{\dimexpr\paperwidth-1.2in} 
          \centering
          \colorbox{lightblue}{%
            \parbox{\dimexpr\linewidth-2\fboxsep}{\footnotesize
              © 2025 IEEE. Personal use of this material is permitted. Permission from IEEE must be obtained for all other uses, in any current or future media, including reprinting/republishing this material for advertising or promotional purposes, creating new collective works, for resale or redistribution to servers or lists, or reuse of any copyrighted component of this work in other works.

            This is the accepted version of the article:  \textit{Learning-Enabled Adaptive Power Capping Scheme for Cloud Data Centers}.\\
            This paper has been accepted for publication in IEEE Transactions on Smart Grid, 2025.%
            }}%
        \end{minipage}
      }%
    }%
  }%
}

\begin{document}

\TopNotice
\title{Learning-Enabled Adaptive Power Capping Scheme for Cloud Data Centers}

\author{Yimeng Sun,~\IEEEmembership{Student Member,~IEEE,}
        Zhaohao Ding,~\IEEEmembership{Senior Member,~IEEE,}\\
        Payman Dehghanian,~\IEEEmembership{Senior Member,~IEEE,}
        and Fei Teng,~\IEEEmembership{Senior Member,~IEEE}% <-this % stops a space

\thanks{This work was supported in part by the National Key R\&D Program of China under Grant 2023YFE0119800, and in part by the National Natural Science Foundation of China under Grants 52277095. \emph{(Corresponding author: Zhaohao Ding.)}}
\thanks{Yimeng Sun and Zhaohao Ding are with the State Key Laboratory of Alternate Electrical Power System with Renewable Energy Sources, North China Electric Power University, Beijing 102206, China (e-mail: yimeng.sun@ncepu.edu.cn; zhaohao.ding@ncepu.edu.cn).}
\thanks{Payman Dehghanian is with the Department of Electrical and Computer Engineering, The George Washington University, Washington, DC 20052 USA (e-mail: payman@gwu.edu).}
\thanks{Fei Teng is with the Department of Electrical and Electronic Engineering, Imperial College London, SW7 2BU, London, U.K. (e-mail: f.teng@imperial.ac.uk).}}

\maketitle

\begin{abstract}
The rapid growth of the digital economy and artificial intelligence has transformed cloud data centers into essential infrastructure with substantial energy consumption and carbon emission, necessitating effective energy management. However, existing methods face challenges such as incomplete information, uncertain parameters, and dynamic environments, which hinder their real-world implementation. This paper proposes an adaptive power capping framework tailored to cloud data centers. By dynamically setting the energy consumption upper bound, the power load of data centers can be reshaped to align with the electricity price or other market signals. To this end, we formulate the power capping problem as a partially observable Markov decision process. Subsequently, we develop an uncertainty-aware model-based reinforcement learning (MBRL) method to perceive the cloud data center operational environment and optimize power-capping decisions. By incorporating a two-stage uncertainty-aware optimization algorithm into the MBRL, we improve its adaptability to the ever-changing environment. Additionally, we derive the optimality gap of the proposed scheme under finite iterations, ensuring effective decisions under complex and uncertain scenarios. The numerical experiments validate the effectiveness of the proposed method using a cloud data center operational environment simulator built on real-world production traces from Alibaba, which demonstrates its potential as an efficient energy management solution for cloud data centers.
\end{abstract}


\IEEEpeerreviewmaketitle

\section{Introduction}

\IEEEPARstart{W}{ith} the rapid advancement of the digital economy and artificial intelligence, cloud data centers have become essential infrastructure in modern society. However, this growth has led to a substantial increase in energy consumption and carbon emissions, creating a critical need for effective energy management to reduce operation cost and promote global environmental sustainability.

Currently, mainstream energy management methods for cloud data centers can be categorized into two lines: job scheduling and resource management. Job scheduling approaches focus on optimizing job allocation and execution to reshape energy consumption. 
These methods typically formalize the scheduling process as bin-packing problems, incorporating factors such as environmental dynamics, job and server heterogeneity, and execution order dependencies \cite{li2019deepjs,cambazard2013bin,bahrami2018data,mao2019learning,liu2023online}. By leveraging machine learning or mathematical programming techniques, these approaches effectively reduce energy costs in data centers while responding to the flexible operation needs of the power grid \cite{li2014modeling,lin2020two,cao2024managing,zhang2019data,ding2018emission}. Alternatively, resource management approaches optimize energy consumption through dynamic scaling or reallocation of resources across hardware and virtual resource levels. At the hardware level, the dynamic voltage and frequency scaling technique is widely adopted by dynamically adjusting the chip voltage and frequency based on job execution characteristics to reduce chip-level energy consumption and overall power usage \cite{masoudi2021energy,shojafar2016energy,toor2019energy,yang2024energy}. At the virtual resource level, resource management involves scaling resource allocation limits and dynamically migrating or reallocating containers and virtual machines \cite{mao2016resource,khodaverdian2024energy,rezakhani2024energy,ma2023virtual,beloglazov2012energy}. These methods enable precise allocation and fine-tuning of computing resources for individual jobs by analyzing their resource needs and service level agreements (SLAs).

Despite their great potential, the real-world implementation of the aforementioned methods is hindered by incomplete information constraints induced by both external and internal reasons. Externally, privacy restrictions between end-user and data center may limit the data center operator's capability to to schedule and control those computing jobs from the cloud tenants \cite{del2016survey}. Internally, the management structures of large cloud companies create multiple divisions based on their major responsibility, such as the divisions responsible for energy management (i.e., energy managers) and job scheduling (i.e., job schedulers) \cite{gallagher2025datacenteraccess,shehabi2016united}. Information exchange and data access barriers exist between different divisions due to security compliance requirements. These barriers could prevent the energy manager, who is in charge of optimizing the power demand, from effectively controlling or even perceiving the job-level operations, thereby complicating the implementation of job-centric energy management schemes in cloud data centers.

To overcome the challenges posed by incomplete information, power capping technique has emerged as a macro-level approach that eliminates the need for detailed job-level information. The major cloud data center operators, such as Google \cite{radovanovic2022carbon}, Microsoft \cite{kumbhare2021prediction}, and Tencent \cite{wu2018precise} have explored power capping techniques to regulate overall power consumption patterns. Researchers have also investigated cluster-level power capping methods. Wang \textit{et al.} \cite{wang2017dynamic} propose an application-aware power scheduling policy, which dynamically allocates power budgets based on application-specific power performance profiles. Ramesh \textit{et al.} \cite{ramesh2019understanding} present a power capping approach that precisely assesses the impact of power capping on computation performance degradation. Petoumenos \textit{et al.} \cite{petoumenos2015power} develop a power capping policy based on job arrival predictions, aiming to achieve a balance between performance loss and energy savings. Wu \textit{et al.} \cite{wu2013classified} propose a classified power capping framework using distribution trees to optimize workload power allocation based on SLAs and power characteristics.However, these methods are often limited by a reliance on one-shot optimization and predefined rules, which fail to account for the dynamic and complex nature of cloud data center operations. As a result, they often lead to suboptimal power capping decisions, causing either excessive energy consumption or degraded service quality. 

To effectively manage energy consumption through power capping, it is essential to understand how the power capping decisions influence computing resource management and job execution processes. In other words, the job scheduler may not respond as expected when faced with a power capping constraint. Thus, inaccurate modeling of the response pattern could hinder the capability of cloud data center to achieve the desired power load curve and Quality of Service (QoS) via the power capping decisions \cite{zhunquejianmo}. However, the internal job scheduling strategy is often inaccessible to energy managers due to privacy and compliance restrictions, which makes the direct modeling of scheduler behavior a significant challenge. Moreover, the dynamic nature of cloud data center operations, characterized by constantly changing scheduling policies and uncertain operational boundaries, further complicates accurate modeling and effective decision-making \cite{complex1,complex2}. Given these challenges, reinforcement learning (RL) has emerged as a promising framework as it does not rely on an accurate physical model. Instead, RL can learn response patterns and incorporate them into the decision-making process by interacting with the dynamic environment \cite{li2017deep}. Although RL has not yet been applied to power capping decisions, it has shown strong potential in related fields, such as job scheduling and resource management in data centers \cite{mao2016resource,mao2019learning,liu2023online,khodaverdian2024energy,rezakhani2024energy}. However, the commonly adopted model-free structure of these RL methods typically requires extensive interactions with the real-world environment, which could be impractical for cloud data centers due to the high cost of exploration \cite{mfrlcon,mfrlcon2}. That is to say, the learning process often involves intermediate policies that may lead to temporarily inefficient states, making excessive real-world interactions prohibitively expensive. Therefore, efficiently learning the response behavior and making power capping decisions with few interactions in such a dynamic and uncertain cloud data center environment remains an open area and requires further investigation.

To address the aforementioned challenges, we propose an adaptive power capping scheme to optimize energy management decisions of cloud data centers. We construct a model-based reinforcement learning (MBRL) framework to characterize the impact of power capping decisions on cloud data center operation, which enable fast learning of effective power capping policy and reducing the cost of multi-round interactions. Additionally, we address the challenges posed by the highly dynamic cloud computing environments via a two-stage uncertainty-aware optimization algorithm. The main contributions of this paper are summarized as follows.

1) We propose an adaptive power capping framework tailored to cloud data center energy management, which can operate without requiring access to job-level details. It aligns with the privacy-preserving and security compliance requirements, making it suitable for practical industrial implementation in cloud data centers.

2) We construct an MBRL model for fast learning of power capping decisions. It reduces the reliance on costly interactions with environment through learning an approximate model of the transition dynamics. Also, we prove the optimality gap of the proposed model under finite iteration rounds.

3) We develop a two-stage uncertainty-aware optimization algorithm to address the associated uncertainties. By integrating uncertainty into both the environment modeling and decision-making processes, we improve the adaptability of power capping policy under stochastic environment.

The remainder of this paper is organized as follows. Section II elaborates the energy management in cloud data centers through power capping and formulates it as a partially observable Markov decision process (POMDP). Section III introduces the MBRL solution for optimizing power caps in uncertain environments. Section IV presents the results of case studies, and Section V concludes the paper.

\section{Problem Formulation}
This section elaborates on the problem of optimizing energy consumption in cloud data centers through adaptive power capping and models the problem as a POMDP in the context of MBRL.

\subsection{Problem Description}

\begin{figure}[h]
    \centering
    \includegraphics[width=7.5cm]{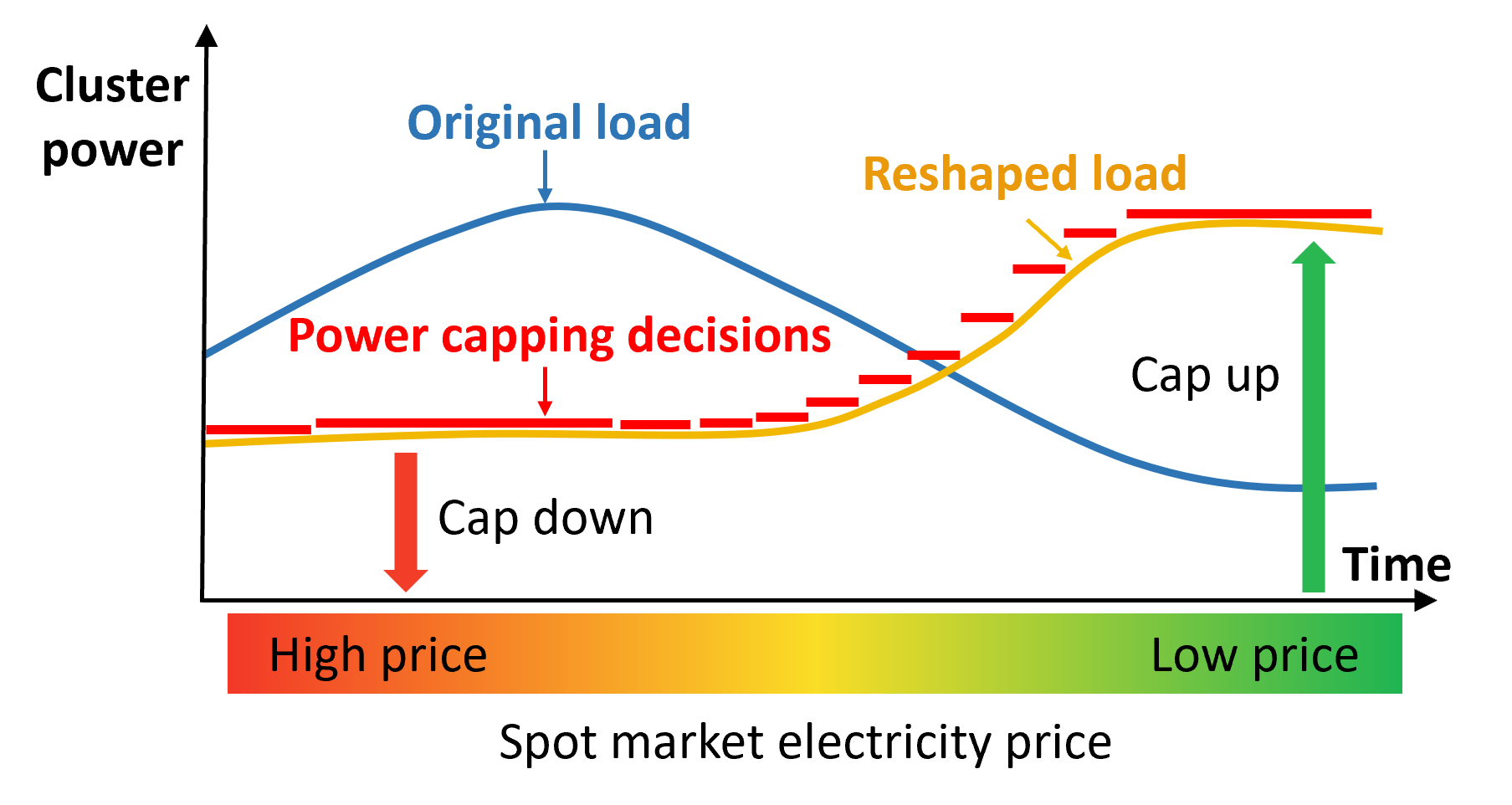}
    \caption{Cloud data center energy management via adaptive power capping.}
    \label{power capping}
    \vspace{-3mm}
\end{figure}

We investigate the problem of adaptive power capping in cloud data centers, as illustrated in Fig. \ref{power capping}. The energy manager dynamically adjusts the power cap (i.e., the energy consumption upper bound) to align with varying electricity price signals, lowering the cap during periods of high electricity prices and raising it when prices are low. This adjustment reshapes the power load profile of the cloud data center, enabling more effective responses to the varied electricity prices and reducing energy costs. It should be noted that this incentive signal can be in other forms, such as marginal carbon intensity or contracted renewable generation profile, to achieve alternative energy management objectives.

\begin{figure}[h]
    \centering
    \includegraphics[width=7cm]{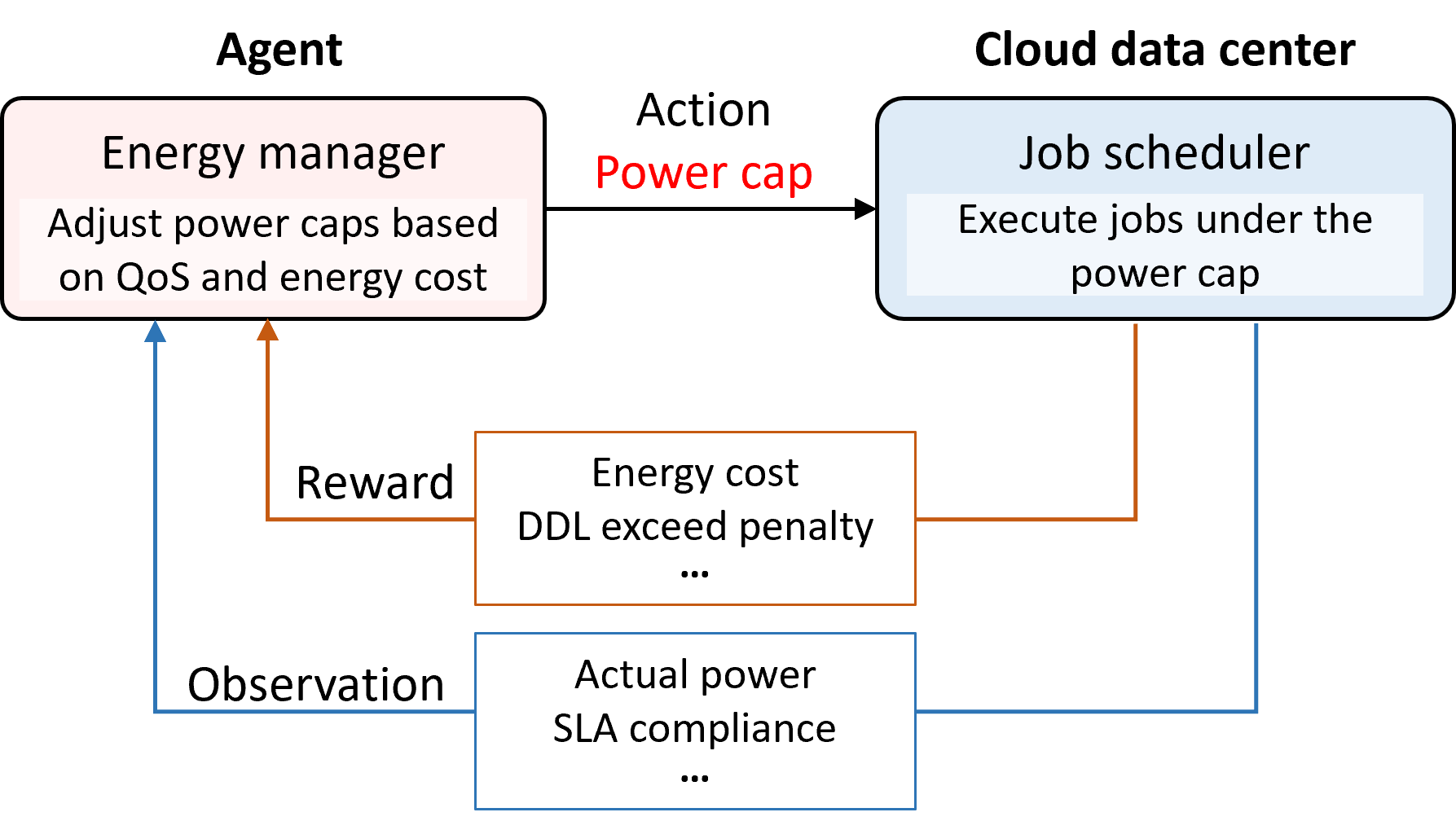}
    \caption{Interactions between energy manager and job scheduler.}
    \label{interaction}
\end{figure}

We assume that the cloud data center acts as a price taker, receiving the locational marginal price from the distribution grid and adjusting its power cap accordingly to optimize energy consumption costs. However, in real-world implementation, power capping decisions account not only for electricity prices but also for the execution of jobs within the cloud data center to ensure QoS.As previously discussed, the energy manager faces the challenge of incomplete information constraints, due to the lack of perception of detailed job-level information, such as execution logic, resource utilization, or job deadlines (DDLs). Consequently, the energy manager must rely on feedback from interactions with the job scheduler to support power capping decisions, as shown in Fig. \ref{interaction}.

\vspace{-3mm}

Specifically, the energy manager determines power caps for the cluster based on electricity prices and feedback from job scheduling process. The job scheduling process then receives the power cap and treats it as an upper bound constraint, guiding the job execution and resource allocation decisions. With the updated operation constraint, the job scheduling process ensures that the power demand of computing resources does not violate the specified upper bound. However, it should be noted that an overly conservative power cap may narrow the operating boundary, potentially leading to increased job delays and SLA violations. As jobs are executed, the job scheduling process provides feedback to the energy manager on the resulting power load and necessary cloud operation information, such as SLA compliance. Subsequently, the energy manager dynamically adjusts the power capping decisions based on the feedback information to optimize energy consumption costs while maintaining the required QoS.

However, the aforementioned power capping process faces challenges due to inherent uncertainties and dynamics in the data center operation environment. On the one hand, the high volatility of electricity prices and the unpredictable job arrival patterns introduce uncertainties that complicate the power capping decision-making. On the other hand, the ever-changing job scheduling policy adds another layer of unpredictability, as different scheduling algorithms may produce diverse outcomes even when constrained by the same power cap. Furthermore, these uncertainties and dynamics necessitate extensive learning interactions to achieve an effective capping policy. The large number of interaction iterations can make the learning process costly, as the intermediate policy may temporarily place the data center in an inefficient operation state. To address these challenges, we construct the following POMDP in the context of MBRL. 

\subsection{POMDP Formulation}

We model the energy manager as an agent that interacts with the uncertain environment to gather operation information and make adaptive power capping decisions. We formulate the power capping process as a POMDP denoted by the tuple $\Gamma = \langle S, A, P, R, \gamma \rangle$.

\textbf{\textit{State:}} The state variable $s_t \in \mathcal{S}_t$ for the energy manager in cloud data centers consists of three components, which are the cluster power information $s^{cluster}_t$, the job scheduler response information $s^{job}_t$, and the price information $s^{pr}_t$. $s^{cluster}_t$ is directly observable by the energy manager, including the current time step $t$, the power cap $P^{cap}_t$, and the actual power load $P^{a}_t$. In contrast, $s^{job}_t$ is obtained through interactions with the job scheduler. It contains the aggregated SLA violation status $SLA^{vio}_t$ and total computing resource requirements $res_t$, which are calculated as follows:
\begin{equation}
    SLA^{vio}_t = \sum_{n=1}^{N} \mathbb{I}_{t > j_n^{ddl}} (t - j_n^{ddl}),
\end{equation}
\begin{equation}
    res_t = \sum_{n=1}^{N} \mathbb{I}_{res_n^{un} > 0} res_n^{un},
\end{equation}
where $\mathbb{I}_{t > j_n^{ddl}}$ is an indicator function that takes 1 if the current time $t$ exceeds the job $n$'s deadline $j_n^{ddl}$, and 0 otherwise. Similarly, $\mathbb{I}_{res_n^{un} > 0}$ takes 1 if job $n$ has unfulfilled resource demands $res_n^{un}$, and 0 otherwise. It should be noted that $SLA^{vio}_t$ and $res_t$ represent aggregated information provided to the energy manager in accordance with privacy and security compliance requirements, preventing the inference of job-level details. $s^{env}_t$ is the spot market electricity price $p_t$ at time $t$.

\textbf{\textit{Action:}} The energy manager's action $a_t \in \mathcal{A}_t$
is to set power cap $P^{cap}_t$ for the cluster. $a_t$ is chosen from the discrete values in the set $[0,1\%,2\%,...100\%]$.

\textbf{\textit{Transition Function:}} The transition function is defined as ${{\mathcal{P}}(s_{t + 1} \mid s_{t}, a_{t}) : \mathcal{S}_t \times \mathcal{A}_t \to \mathcal{S}_{t+1}}$. In the context of MBRL, we establish the environment model network \( \hat{\mathcal{P}}_\theta(s_{t + 1} \mid s_{t}, a_{t}) \) to approximate the job scheduling process taking the power capping decisions as boundary conditions. \(\hat{\mathcal{P}}_\theta(s_{t + 1} \mid s_{t}, a_{t})\) represents the probability of transitioning to state \( s_{t+1} \) after taking action \( a_t \) in state \( s_t \). It characterizes the impact of power capping decisions on environment transitions. The specific construction and learning process of \( \hat{\mathcal{P}}_\theta(s_{t + 1} \mid s_{t}, a_{t}) \) is detailed in Section III.

\textbf{\textit{Reward and $\gamma$:}} The reward function for the energy manager accounts for energy consumption costs, SLA violation penalties, and power cap adjustment costs. It is defined as:
\begin{equation}
    r_t = r^{energy}_t + r^{ddl}_t + r^{cap}_t.
\end{equation}
The components are calculated as follows.
\begin{itemize}
\item $r^{energy}_t$: The energy consumption cost at time $t$ is defined as the negative product of the actual power consumption $P^{a}_t$ and the electricity price $p_t$.
\begin{equation} 
r^{energy}_t = -P^{a}_t \times p_t. 
\end{equation} 
\item $r^{ddl}_t$: If a job fails to meet the specified deadline according to the SLA, a penalty is incurred. The deadline violation reward is calculated as the total time exceeded at time $t$, multiplied by a large penalty:
\begin{equation} 
r^{ddl}_t = - SLA^{vio}_t \times pen, 
\end{equation}
where ${pen}$ is the penalty set by the SLA for exceeding deadlines.
\item $r^{cap}_t$: A wide range of power cap adjustments can affect job execution and lead to wear and tear of hardware \cite{deng2014reliability}. To minimize the impact of the adjustment, we define the adjustment cost as follows:
\begin{equation} 
r^{cap}_t = -(P^{max} - P^{cap}_t) \times adj_c, \end{equation}
where $P^{max}$ represents the maximum rated power of the cluster, and $adj_c$ is the cost associated with the adjustment.
 \end{itemize}
 
Based on the aforementioned single-step reward $r_t$, the energy manager makes decisions according to the long-term cumulative discounted return $R_t = r_t + \gamma r_{t+1} + \cdots + \gamma^{T - 1 - t}r_T$. $\gamma \in [0,1]$ is the discount factor that balances the importance of immediate versus future rewards.

\textbf{\textit{Policy:}} $\pi ({a_t}{\left|s_t\right.})$ specifies the probability of taking action $a_t$ in state $s_t$. The goal of the adaptive power capping problem is to find an optimal policy ${\pi^*}$ that enables the best trade-off between energy consumption costs and job execution. The objective function can be formulated as:
 \begin{equation}
 {\pi ^*}(\left. {{a_{t}}} \right|{s_{t}}) = \mathop {\arg \max }\limits_{{a_{t}} \in {{{\mathcal{A}}}_{t}}} Q_\phi({s_{t}},{a_{t}}).
\end{equation}
where $Q_\phi({s_{t}},{a_{t}})$ is the distributional state-action value function. It is updated  based on the Bellman equation:
\begin{equation}
    \mathcal{T}^* Q_\phi(s_t, a_t) = r_t + \gamma \sum_{s_{t+1}} \hat{P}_\theta(\hat{s}_{t+1}|s_t, a_t) \max_{a_{t+1}} Q_\phi(s_{t+1}, a_{t+1}).
\end{equation}
Here, $\mathcal{T}^*$ is the optimal Bellman operator that reflects the optimal long-term cumulative reward when following the best possible policy. $\hat{P}_\theta(s_{t+1}|s_t, a_t)$ is the transition probability accounting for uncertainties in the environment.

\section{Solution Method}
To address the proposed POMDP, we develop an uncertainty-aware MBRL scheme to achieve adaptive power capping decisions. Specifically, we first present an MBRL framework that facilitates the fast learning of adaptive power capping policy in dynamically changing environments. Within this framework, we then introduce a two-stage uncertainty-aware algorithm to capture uncertainties in cloud data centers.Accordingly, we derive the optimality gap of the proposed scheme under finite iterations, offering a theoretical guarantee for its decision-making performance in cloud data centers.

\vspace{-3mm}
\subsection{Structure of the MBRL Framework}
\begin{figure}[h]
    \centering
    \includegraphics[width=8.8cm]{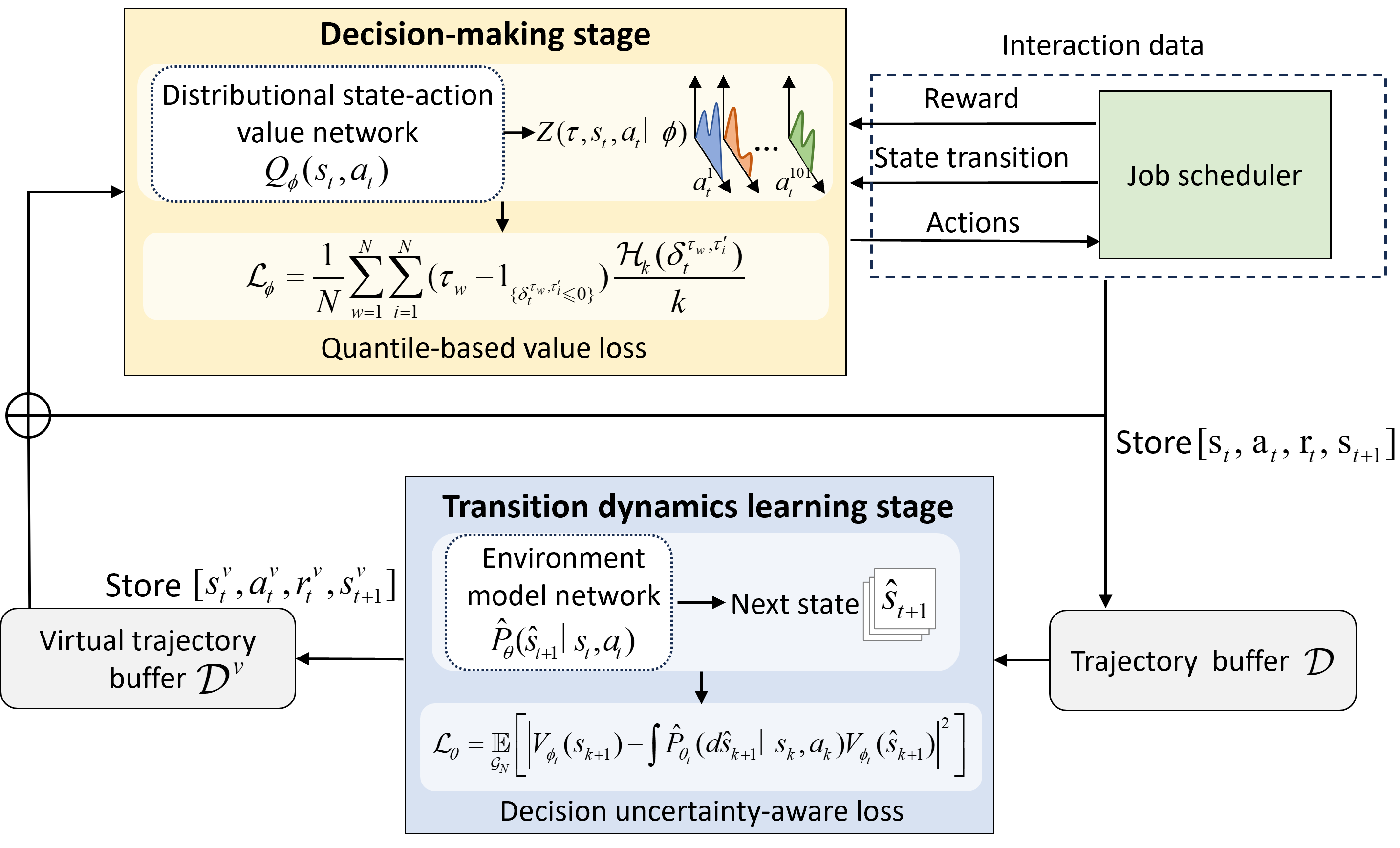}
    \caption{Workflow of the uncertainty-aware MBRL scheme.}
    \label{MBRL}
    \vspace{-8mm}
\end{figure}

The MBRL framework consists of two components: the environment model network \( \hat{P}_\theta(\hat{s}_{t+1} \mid s_t, a_t) \) that learns transition dynamics of the job scheduling process, and the distributional state-action value network \( Q_\phi(s_t, a_t) \) that handles decision-making of energy manager, as shown in Fig. \ref{MBRL}.

The energy manager selects power capping actions based on the output of \( Q_\phi(s_t, a_t) \) according to (7), then interacts with the job scheduler to receive critical operation information on the job scheduling process. These interactions generate trajectories \( g_t = [s_t, a_t, r_t, s_{t+1}] \), which are stored in the buffer \( \mathcal{D} \). Subsequently, the trajectories in \( \mathcal{D} \) are used to train \( \hat{\mathcal{P}}_\theta(\hat{s}_{t+1} \mid s_t, a_t) \) in a supervised manner to learn the dynamic responses of the job scheduler to power capping actions. Training continues until the loss decreases to a preset threshold \(\delta\), at which point \( \hat{\mathcal{P}}_\theta(\hat{s}_{t+1} \mid s_t, a_t) \) is considered capable of effectively simulating transition dynamics. Next, we use the trained \( \hat{\mathcal{P}}_\theta \) and \( Q_\phi \) to generate virtual trajectories \( g_t^v = [s_t^v, a_t^v, r_t^v, s_{t+1}^v] \), which are stored in the virtual buffer \( \mathcal{D}^v \). The combined buffers \( \mathcal{D}^v \) and \( \mathcal{D} \) are then used to further train the decision network \( Q_\phi \), improving the speed and effectiveness of learning the adaptive power capping policy. By utilizing virtual trajectories generated by the environmental model \( \hat{P}_\theta(\hat{s}_{t+1} \mid s_t, a_t) \), the policy learning process can be accelerated, enabling the energy manager to achieve fast learning with fewer interactions with the cloud data center's operational environment. However, to balance computational cost and learning efficiency, we constrain both the rollout horizon and the batch size for virtual trajectory generation, ensuring that the process remains computationally manageable.

\vspace{-3mm}

\subsection{Two-Stage Uncertainty-Aware Optimization}
This subsection proposes a two-stage uncertainty-aware algorithm, which consists of a transition dynamics learning stage and a decision-making stage, to effectively capture uncertainties in the cloud data center environment.

In the decision-making stage, the distributional state-action value network \( Q_\phi \) is constructed via a quantile regression approach. The network takes \( s_t \), \( a_t \), and sampled quantiles \( \tau_s = [\tau_1, \ldots, \tau_N] \) as inputs and outputs the state-action value distribution \( Z(\tau, s_t, a_t \mid \phi) \) according to the input quantiles. Subsequently, the state-action value is computed as:
\begin{equation}
   Q_{\phi}(s_t, a_t) = \mathop\mathbb{E}\limits_{\tau  \in {\tau _s}} [Z(\tau, s_t, a_t \mid \phi)].
\end{equation}
By representing the value distribution through quantiles, the full distributional information of the returns is captured, enabling the energy manager to account for uncertainties in decision-making rather than relying solely on the expected value. Given the superior performance of the Wasserstein distance in measurement accuracy and training convergence speed \cite{dabney2018implicit}, we perform quantile regression by minimizing the Wasserstein distance between the online and target distributions corresponding to the sampled quantiles. Specifically, the quantile regression is performed by minimizing the Wasserstein distance using the Huber loss ${{{\mathcal{H}}}_k}$ as the quantile regression loss function. The loss function is defined as:
\begin{equation}\label{}
{{\mathcal{L}_\phi}} = \frac{1}{{N}}\sum\limits_{w = 1}^{{N}} {\sum\limits_{i = 1}^{{N}} {({\tau _w} - {1_{\{ \delta _{t}^{{\tau _w},{{\tau '}_i}} \leqslant 0\} }})} } \frac{{{{{\mathcal{H}}}_k}(\delta _{t}^{{\tau _w},{{\tau '}_i}})}}{k},
\end{equation}
\begin{equation}
    \delta _{t}^{\tau ,\tau '} = r_t + \gamma Z(\tau ', s_{t+1}, a_{t+1}) - Z(\tau, s_t, a_t),
\end{equation}
where \emph{k} is the threshold value of the Huber loss function, $N$ is the number of quantile samples, $\tau $ and $\tau '$ are sampled quantiles from $U[0,1]$ for online distribution and target distribution, respectively. This allows incorporating the uncertainties of the cloud data center environment into the decision-making process through the quantile regression approach.

In the transition dynamics learning stage, an environment model \( \hat{P}_\theta(\hat{s}_{t+1} \mid s_t, a_t) \) is constructed to embed decision process knowledge and capture feedback uncertainties. Trajectories ${\mathcal{G}_N} \subset \mathcal{D}$ are used to predict the possible next state \( \hat{s}_{t+1} \) after taking action $a_t$ in state $s_t$, i.e., learning the mapping ${\mathcal {S}_t} \times {\mathcal {A}_t} \to {\mathcal{S}_{t+1}}$. To account for uncertainty in the energy manager's decision-making process during learning of the environment model, a decision uncertainty-aware loss function is defined as:
\begin{equation}
\mathcal{L}_\theta = \mathop\mathbb{E}\limits_{\mathcal{G}_N} \left[ \left| V_{\phi_t}(s_{k+1}) - \int \hat{P}_{\theta_t}(d\hat{s}_{k+1} \mid s_k, a_k) V_{\phi_t}(\hat{s}_{k+1}) \right|^2 \right],
\end{equation}
where \( \mathcal{G}_N \) represents a set of \( N \) sampled trajectories \([s_k, a_k, s_{k+1}]\). The state value function \( V_{\phi_t}(s_t) \), which represents the expected return of state \( s_t \), is defined as:
\begin{align}
V_{\phi_t}(s_t) &= \mathop\mathbb{E}\limits_{a_t \sim \pi(\cdot \mid s_t)} \left[ Q_{\phi_t}(s_t, a_t) \right] \notag\\
&= \mathop\mathbb{E}\limits_{a_t \sim \pi(\cdot \mid s_t)} \left[ \mathop\mathbb{E}\limits_{\tau \in \tau_s} \left[ Z(\tau, s_t, a_t \mid \phi_t) \right] \right].
\end{align}

By utilizing the distribution information $Z(\tau, s_t, a_t \mid \phi_t)$ from the decision-making process, the state-value function \( V_{\phi_t}(s_t) \) embeds the structure and knowledge of the uncertainty of the decision problem into the model learning process. This approach captures both the inherent uncertainties in the cloud data center environment and information relevant to decision-making optimization. In contrast, loss functions such as MSE, $L_1$ loss, or other probability-based metrics primarily focus on prediction accuracy, often neglecting the intention of the decision (i.e., optimization direction) and the underlying uncertainties in the decision-making process. By effectively capturing these environmental uncertainties, the decision uncertainty-aware loss function aligns closely with the decision-making process, aiding the energy manager in better exploring the feedback of the job scheduler and improving the decision-making efficiency and adaptability.

\vspace{-3mm}
\subsection{Optimality Gap under Finite Iterations}

The method proposed in Sections III-A and B enables fast learning of the power capping policy, reducing time and costs associated with the intermediate interactions in cloud data centers. This subsection derives the optimality gap \( Z^{\text{gap}} = \mathcal{W}(Z^*, Z^{\pi_K}) \) between the distribution of the state-action value function \( Z^{\pi_K} = Z(\tau, s_t, \pi_K(s_t)) \), induced by policy \( \pi_K \), and the ideal optimal distribution \( Z^*(\tau, s_t, \pi^*(s_t))\) under finite \(K\) sampling, where \( \mathcal{W}(X, Y) \) denotes the Wasserstein metric. This represents the performance difference between the policy $\pi_K$ and the optimal decision under finite iterations. 

Firstly, we introduce the following definition of state-action value distribution concentrability.

\textbf{Definition 1(Concentrability Coefficient)}: For an arbitrary policy sequence \( (\pi_i)_{i=1}^k \) and the true distribution \( z_k \), \( z_0 \hat{\mathcal{P}}^{\pi_1} \dots \hat{\mathcal{P}}^{\pi_k} \) denotes the future state-action value distribution when the initial state-action follows \( z_0 \) and the agent follows policies \( \pi_1, \dots, \pi_k \). To simplify the expression, $\hat{\mathcal{P}}^{\pi_k}$ denotes the approximate transition probability under policy $\pi_k$. The concentrability coefficient is defined as
\begin{equation}
\overline{c}_{z_0, z_k}(k) = \sup_{\pi_1, \dots, \pi_k} \left\| \frac{d z_0 \hat{\mathcal{P}}^{\pi_1} \dots \hat{\mathcal{P}}^{\pi_k}}{d z_k} \right\|_{2}.
\end{equation}
Considering a discount factor \( 0 < \gamma < 1 \), the discounted average concentrability coefficient is given by:
\begin{equation}
\overline{C}(z_0, z_k) = (1 - \gamma)^2 \sum_{k \geq 1} k \gamma^{k-1} \overline{c}_{z_0, z_k}(k).
\end{equation}
Based on Definition 1, we present Lemma 1 to provide an error upper bound of the energy manager's decisions.

\textbf{Lemma 1(Optimality gap under finite iterations):} \textit{Consider a finite sequence of state-action value function distributions \( (\hat{Z}_k)_{k=1}^K \). Define the environment modeling error as \( \Delta_k = \mathcal{W}(\hat{Z}_{k+1}, T^*_{\mathcal{P}^*}\hat{Z}_k) \) for \( k = 0, 1, \dots, K - 1 \). Let \( \pi_K \) be the greedy policy with respect to \( \hat{Z}_K \), defined by\(\pi_K(s_t) = \mathop{\arg \max}_{a_t \in \mathcal{A}_t} \mathbb{E}_{\tau \in \tau_s} [Z(\tau, s_t, a_t)]\) for all \( s_t \in \mathcal{S}_t \).}

\textit{Then, the following inequality holds:}
\begin{align}
 Z^{\text{gap}}_{z_0} &\leq 2\gamma \left[ \frac{\overline{C}(z_0,z_K)}{(1 - \gamma)^2} \max_{0 \leq k \leq K-1}\Delta_k + 2 \gamma^K R_{\text{max}} \right],
\end{align}
\textit{where \( R_{\text{max}} \) is the maximum single-step reward, as the defined MDP is \( R_{\text{max}} \)-bounded.}

\textbf{Proof:}
Consider the energy manager's optimal policy $\pi^*$. By Bellman optimality, \( \mathcal{T}^{\pi^*} Z^* = \mathcal{T}^* Z^* = Z^* \). Then, applying the properties of triangle inequality for the Wasserstein metric, we obtain:
\begin{align}
\mathcal{W}(Z^*, \hat{Z}_{k+1}) 
&\leq \mathcal{W}(\mathcal{T}^{\pi^*} Z^*, \mathcal{T}^* \hat{Z}_k) + \mathcal{W}(\mathcal{T}^* \hat{Z}_k, \hat{Z}_{k+1}) \notag \\
&\leq \mathcal{W}(\mathcal{T}^{\pi^*} Z^*, \mathcal{T}^{\pi^*} \hat{Z}_k) + \Delta_k \\
&=\gamma \mathcal{P}^{\pi^*} \mathcal{W}(Z^*, \hat{Z}_k) + \Delta_k \notag.
\end{align}
By induction from (17), we have:
\begin{align}
\mathcal{W}(Z^*, \hat{Z}_K) &\leq \sum_{k=0}^{K-1} \gamma^{K-k-1} (\mathcal{P}^{\pi^*})^{K-k-1} \Delta_k \notag \\
&\quad + \gamma^K (\mathcal{P}^{\pi^*})^K \mathcal{W}(Z^*, z_0)
\end{align}
Likewise, applying the properties of triangle inequality for the Wasserstein metric, $Z^{gap}$ is bounded as:
\begin{align}
    Z^{gap} & \leq \mathcal{W}(\mathcal{T}^{\pi^*} Z^*, \mathcal{T}^{\pi_K} \hat{Z}_K) + \mathcal{W}(\mathcal{T}^{\pi_K} \hat{Z}_K, \mathcal{T}^{\pi_K} Z^{\pi_K}) \notag \\
    &\leq \gamma \mathcal{P}^{\pi^*} \mathcal{W}(Z^*, \hat{Z}_K) + \gamma \mathcal{P}^{\pi_K} \mathcal{W}(\hat{Z}_K, Z^*) \notag\\
    &\quad + \gamma \mathcal{P}^{\pi_K} \mathcal{W}(Z^*, Z^{\pi_K}) \notag \\
    &= \gamma (\mathcal{P}^{\pi^*}+\mathcal{P}^{\pi_K}) \mathcal{W}(Z^*, \hat{Z}_K) + \gamma \mathcal{P}^{\pi_K} \mathcal{W}(Z^*, Z^{\pi_K}).
\end{align}
By rearranging, we obtain
\begin{align}
    Z^{gap} &\leq \gamma \mathcal{I}_K^{-1} ( \mathcal{P}^{\pi^*} + \mathcal{P}^{\pi_K} ) \mathcal{W}(Z^*, \hat{Z}_K),
\end{align}
where $\mathcal{I}_K^{-1} = \left( \mathbf{I} - \gamma \mathcal{P}^{\pi_K} \right)^{-1}$ and $\mathcal{I}$ is the identify operator. We combine (20) with (18) to get
\begin{align}
 Z^{gap} &\leq \frac{2(1 - \gamma^{K+1})\gamma}{(1 - \gamma)} \left[ \sum_{k=0}^{K-1} \alpha_k A_k \Delta_k + \alpha_K A_K \mathcal{W}(Z^*, z_0) \right],
\end{align}
where \( \alpha_k \) and \( A_k \) are introduced to simplify notation, defined as follows:
\begin{align}
\alpha_k =\begin{cases} 
\frac{(1-\gamma)\gamma^{K-k-1}}{1-\gamma^{K+1}} , & 0 \leq k < K \\
\frac{(1-\gamma)\gamma^{K}}{1-\gamma^{K+1}}, & k = K
\end{cases}
\end{align}

\begin{align}
A_k = \begin{cases} 
\frac{1}{2} \mathcal{I}_K^{-1} \left[ ( \mathcal{P}^{\pi^*} )^{K-k} +  \mathcal{P}^{\pi_K}(\mathcal{P}^{\pi^*})^{K-k-1} \right], & 0 \leq k < K \\
\frac{1}{2} \mathcal{I}_K^{-1} \left[ ( \mathcal{P}^{\pi^*} )^{K+1} + \mathcal{P}^{\pi_K}( \mathcal{P}^{\pi^*})^{K} ) \right], & k = K
\end{cases}
\end{align}

To compute the expected performance difference, we apply the probability distribution \( z_0 \) to both sides. Then, using Jensen's inequality, we move \( z_0 \) inside, yielding:
\begin{align}
Z^{gap}_{z_0} \leq \gamma_K 
\left[ \sum_{k=0}^{K-1} \alpha_k z_0 A_k \Delta_k + \alpha_K z_0 A_K \mathcal{W}(Z^*, z_0) \right],
\end{align}
where $\gamma_K = \frac{2(1 - \gamma^{K+1})\gamma}{(1 - \gamma)}$.

The two terms on the right side of (24) correspond to the expressions related to the learning error of the environment model and the initial distribution distance, respectively. The detailed derivation process for the upper bounds of these two terms is provided in Appendix A and Appendix B, with the results summarized as follows:
\begin{equation}
  \gamma_K \sum_{k=0}^{K-1} \alpha_k z_0 A_k \Delta_k \leq 2\gamma \frac{\overline{C}(z_0, z_K)}{(1 - \gamma)^2} \max_{0 \leq k \leq K-1}\Delta_k  
\end{equation}
and
\begin{equation}
  \gamma_K \alpha_K z_0 A_K \mathcal{W}(Z^*, z_0) \leq 4\gamma^{K+1} R_{\text{max}}.  
\end{equation}
By substituting (25) and (26) into (24), we finally obtain 
\begin{align}
 Z^{\text{gap}}_{z_0} &\leq 2\gamma \left[ \frac{\overline{C}(z_0,z_K)}{(1 - \gamma)^2} \max_{0 \leq k \leq K-1}\Delta_k + 2 \gamma^K R_{\text{max}} \right].
\end{align}
(27) indicates that the gap between the outcomes of $K$ iterations of policy optimizations and the optimal policy result is upper bounded by an error term. This term is influenced by the model learning error $\Delta_k$ and the $\gamma^{K+1}$ decayed $R_{\text{max}}$, both of which are typically small. Consequently, we prove the effectiveness of the proposed decision-making approach with finite $K$ iterations, which offers theoretical guarantees for the performance of the adaptive power capping policy in cloud data centers. The pseudo-code of the proposed uncertainty-aware MBRL algorithm is presented in Algorithm 1.

\begin{algorithm}
    \caption{Uncertainty-aware MBRL for Adaptive Power Capping}
    Initialize $Q_\phi(s_t, a_t)$ and \( \hat{P}_\theta(\hat{s}_{t+1} \mid s_t, a_t) \) with random weights $\phi_0$ and $\theta_0$\;
    Initialize total episodes $N_{\text{ep}}$, trajectory count $N^{tra}$, sample batch size $B$, training batch size $B^T$, model loss threshold $\delta$, validation loss threshold $\delta_v$, and early stopping counter $C_{\text{stop}}$\;
    \For{$\text{episode} = 1$ to $N_{\text{ep}}$}{
        Initialize state $s_0$\;
        \For{$t = 1$ to $T$}{
            Energy manager sets power capping $a_t$ using $\epsilon$-greedy policy based on Eq. (7)\;
            Execute action $a_t$, observe reward $r_t$ and next state $s_{t+1}$\;
            Store transition $[s_t, a_t, r_t, s_{t+1}]$ in $\mathcal{D}$\;
            \If{$N^{tra} \geq B$}{
                Sample $B$ trajectories from $\mathcal{D}$\;
                \If{$\mathcal{D}^v$ is not empty}{
                    Sample $B$ virtual trajectories from $\mathcal{D}^v$\;
                    Append virtual trajectories to the sampled trajectories\;
                }
                Update $Q_\phi(s_t, a_t)$ using Eq. (10) with either $B$ or $2B$ trajectories\;
            }
            \If{$N^{tra} \geq B^T$ and the number of new trajectories equals $B^T$}{
                Sample $B^T$ trajectories from $\mathcal{D}$ and compute model loss $\mathcal{L}_\theta$ on these samples\;
                \While{$\mathcal{L}_\theta \geq \delta$}{
                    Update \( \hat{P}_\theta(\hat{s}_{t+1} \mid s_t, a_t) \) using gradient descent on Eq. (12)\;
                }
                \If{$\mathcal{L}_\text{val} > \mathcal{L}_\text{train}$ for $C_{\text{stop}}$ consecutive epochs}{
                    Terminate training early to prevent overfitting\;
                    \textbf{break}\;
                }}
            }
        }
\end{algorithm}

\section{Case Study}

\subsection{Experimental Setting}
\begin{figure}[h]
    \centering
    \includegraphics[width=8.5cm]{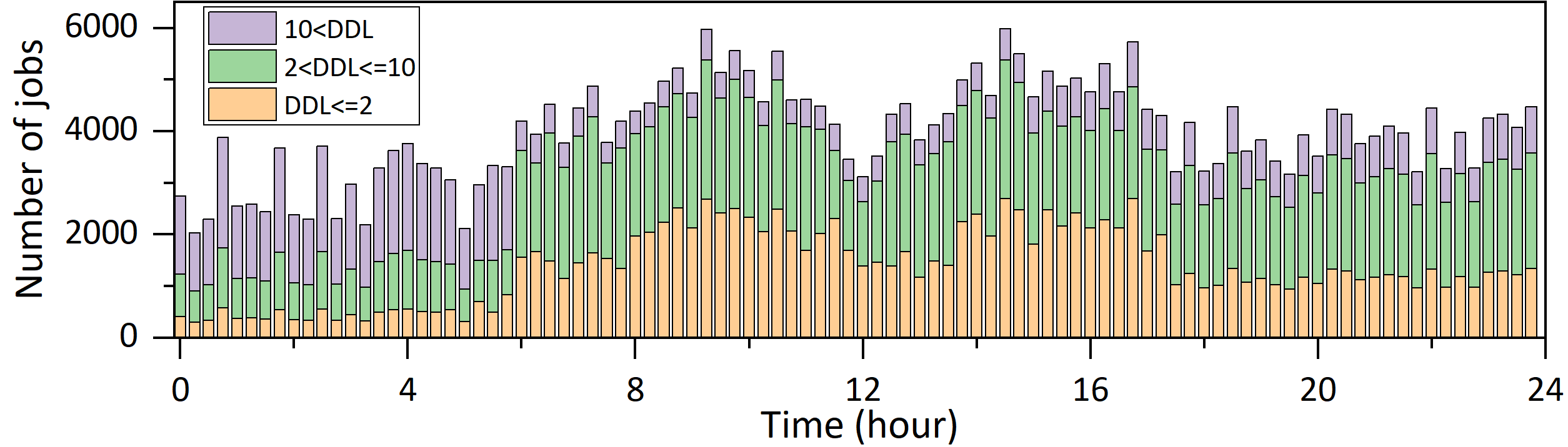}
    \caption{Daily job arrival patterns in cloud data centers.}
    \label{job arrival}
\end{figure}

This section conducts numerical experiments to illustrate the effectiveness of the proposed power capping approach. We mimic the cloud data center operational environment using a simulator based on the computing job scheduling process with power budget constraints (i.e., the power capping decision). We simulate four commonly used real-world job scheduling policies, which are first-come-first-serve (FCFS) \cite{saeed2019load}, round-robin (RR) \cite{chaskar2003fair}, earliest deadline first (EDF) \cite{stankovic1998deadline}, and energy-aware scheduling, to represent varied configuration of cloud data centers. The simulator receives the power capping decisions from the energy manager and then schedules jobs based on the selected scheduling algorithm, providing output on actual power load and critical job execution information to the energy manager. The job arrival pattern is generated based on the real-world production traces of Alibaba \cite{alibaba2017trace}, as shown in Fig. \ref{job arrival}. We model the energy consumption of the job scheduling process using a linear model, based on the resource utilization information of jobs from the real-world data \cite{energymodel}. To account for variations in electricity prices, we use seasonal electricity price data from the PJM market to characterize price fluctuations throughout different seasons \cite{pjm2025market}. Without loss of generality, we select FCFS as the default scheduling policy to interact with the energy manager in the following experiments.

\subsection{Cloud Data Center Energy Management via Adaptive Power Capping}
\begin{figure}[htbp]
    \centering
    \begin{minipage}[b]{\columnwidth}
        \includegraphics[width=\linewidth]{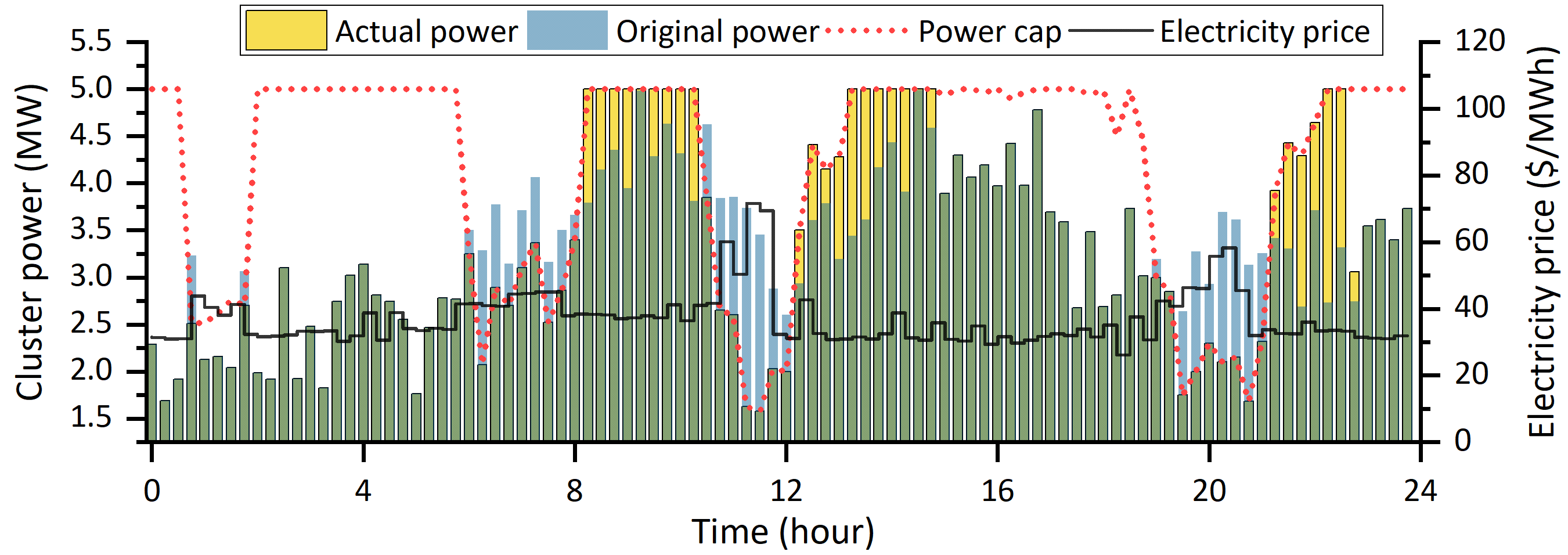}
        \caption*{\small (a)}
        \label{spring}
    \end{minipage}
    \begin{minipage}[b]{\columnwidth}
        \includegraphics[width=\linewidth]{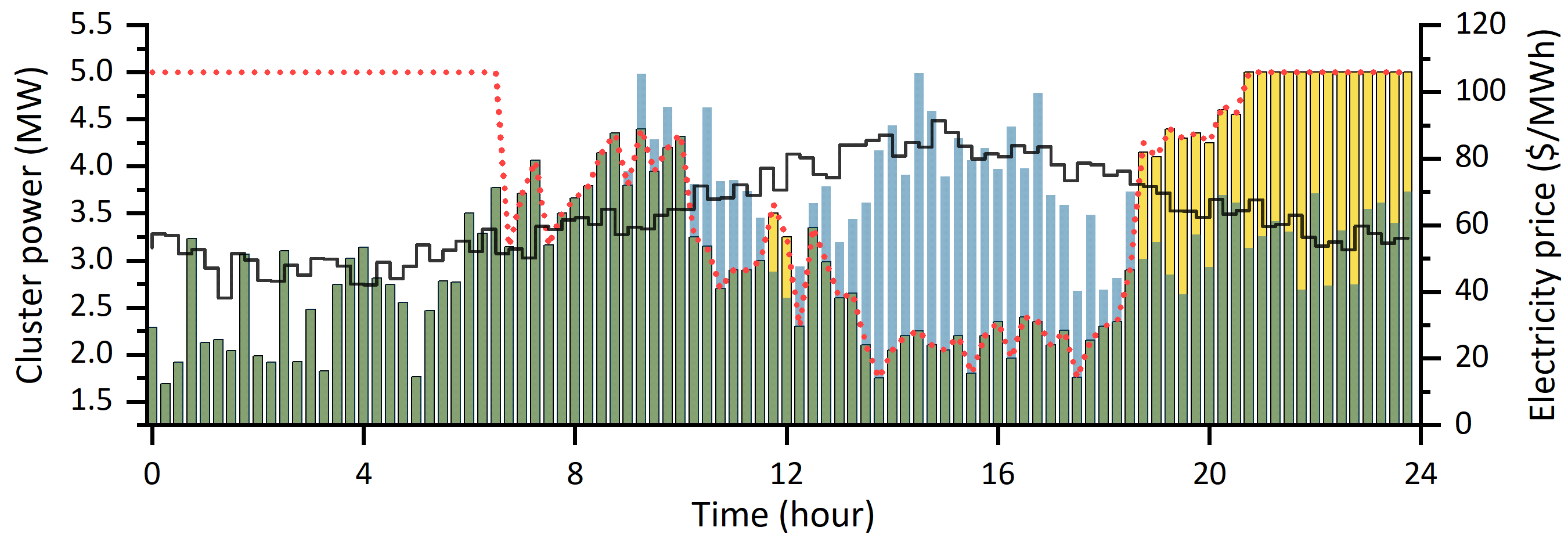}
        \caption*{\small (b)}
        \label{summer}
    \end{minipage}
    \begin{minipage}[b]{\columnwidth}
        \includegraphics[width=\linewidth]{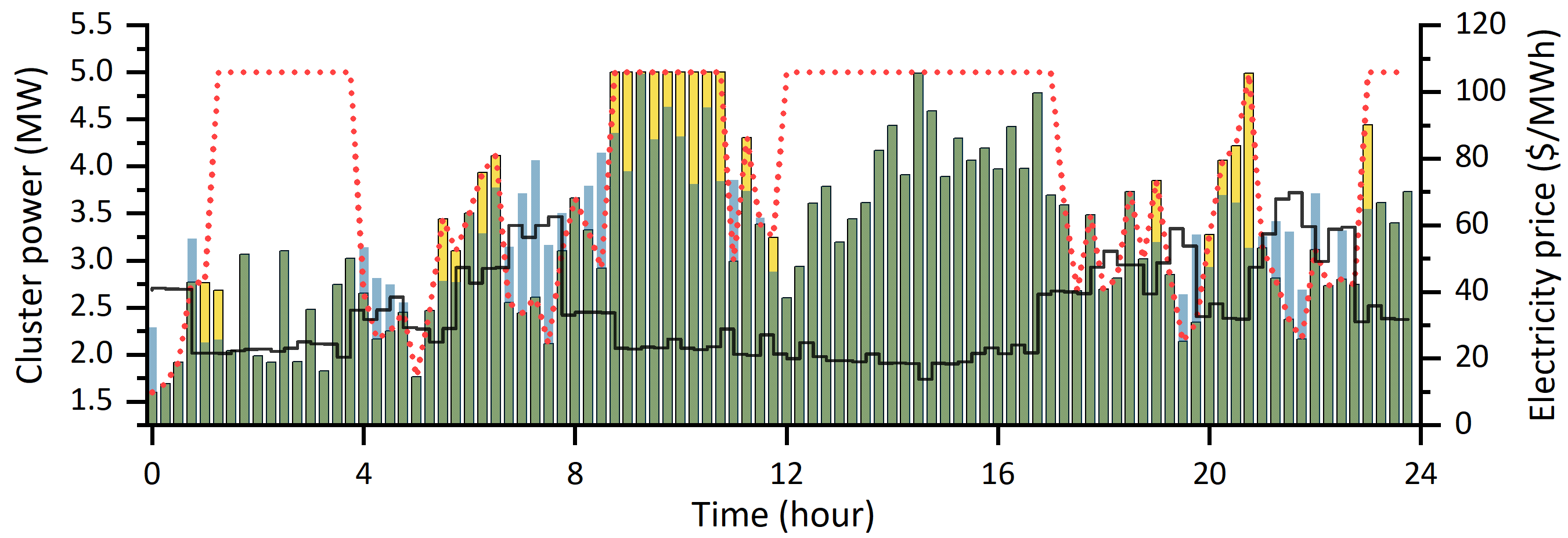}
        \caption*{\small (c)}
        \label{autumn}
    \end{minipage}
    \begin{minipage}[b]{\columnwidth}
        \includegraphics[width=\linewidth]{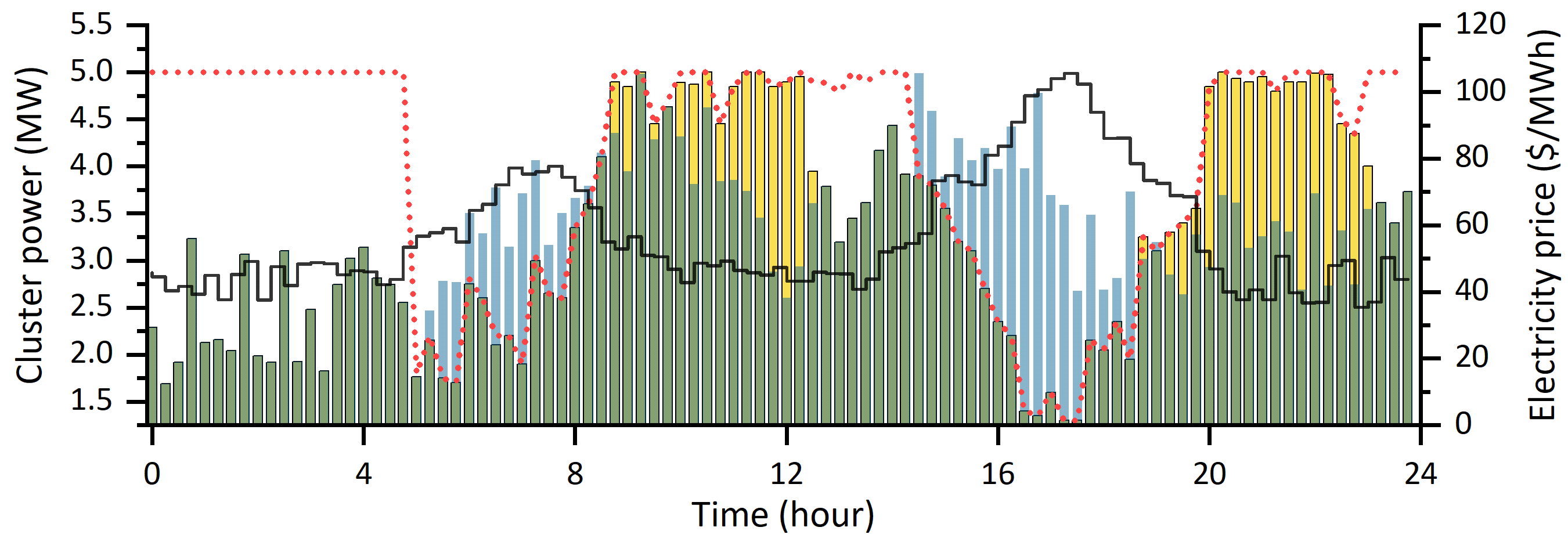}
        \caption*{\small (d)}
        \label{winter}
    \end{minipage}
    \caption{Power capping and actual power results under (a) spring, (b) summer, (c) autumn, and (d) winter scenarios.}
    \label{seasons}
    \vspace{-5mm}
\end{figure}

\begin{figure}[h]
    \centering
    \includegraphics[width=7.8cm]{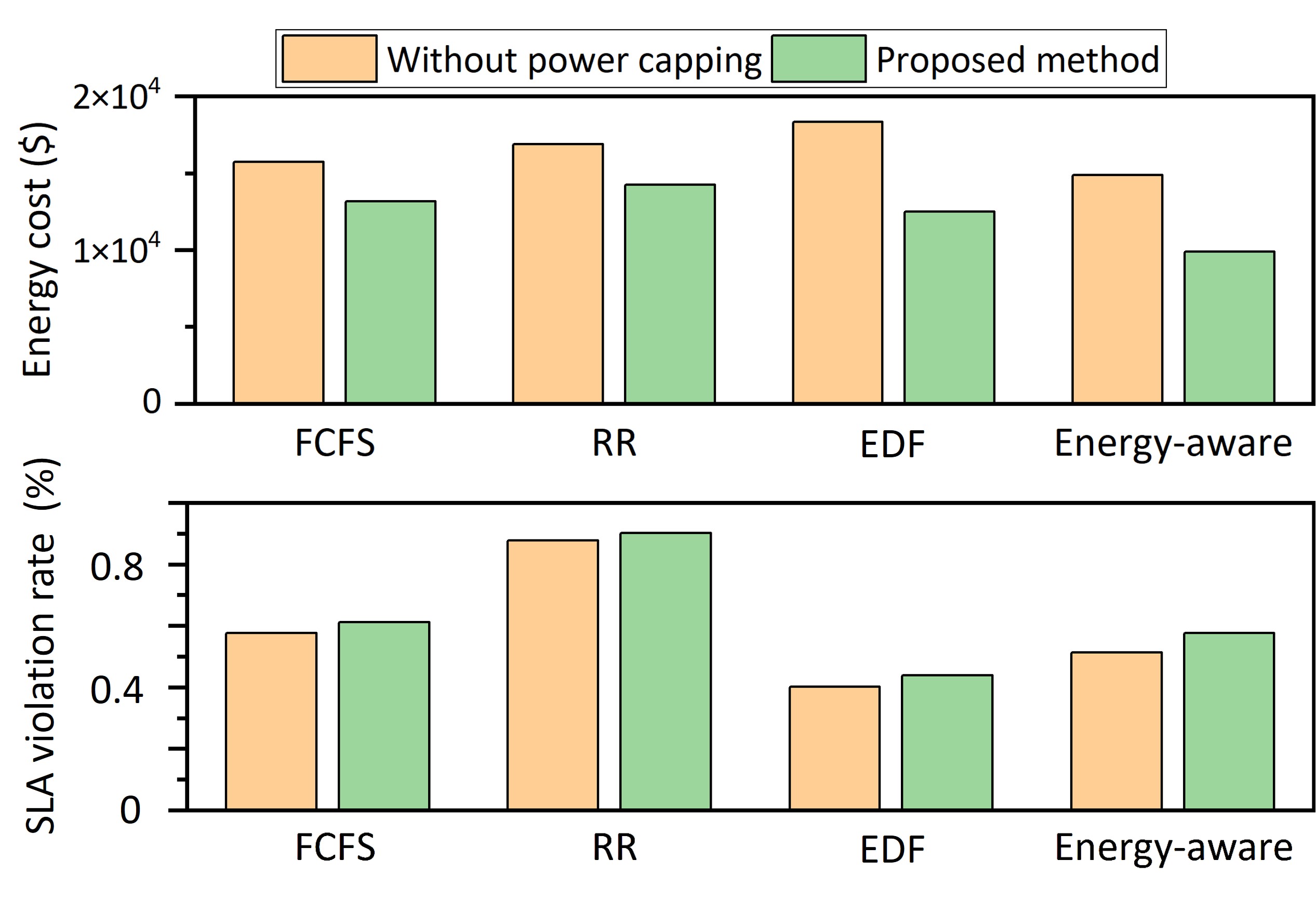}
    \caption{Comparison of energy cost and SLA violation rate with and without power capping across different scheduling policies.}
    \label{different scheduler}
    \vspace{-3mm}
\end{figure}

\begin{figure}[h]
    \centering
    \includegraphics[width=1\linewidth]{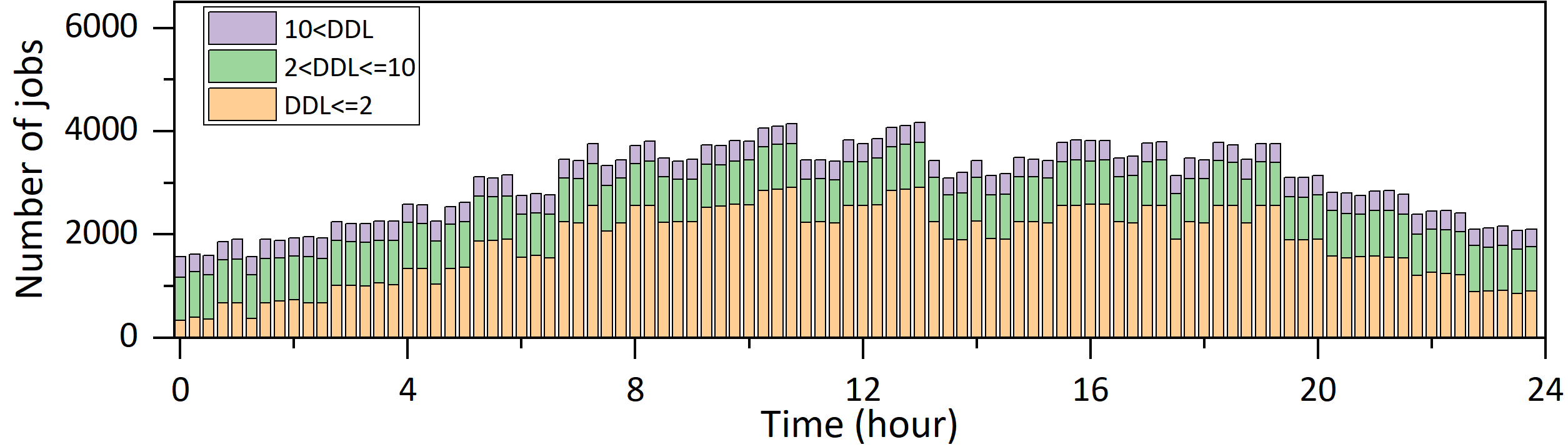}
    \caption{Uniformly distributed online job arrival patterns.}
    \label{fig:arrival_short}
    \vspace{-3mm}
\end{figure}

\begin{figure}[h]
    \centering
    \includegraphics[width=1\linewidth]{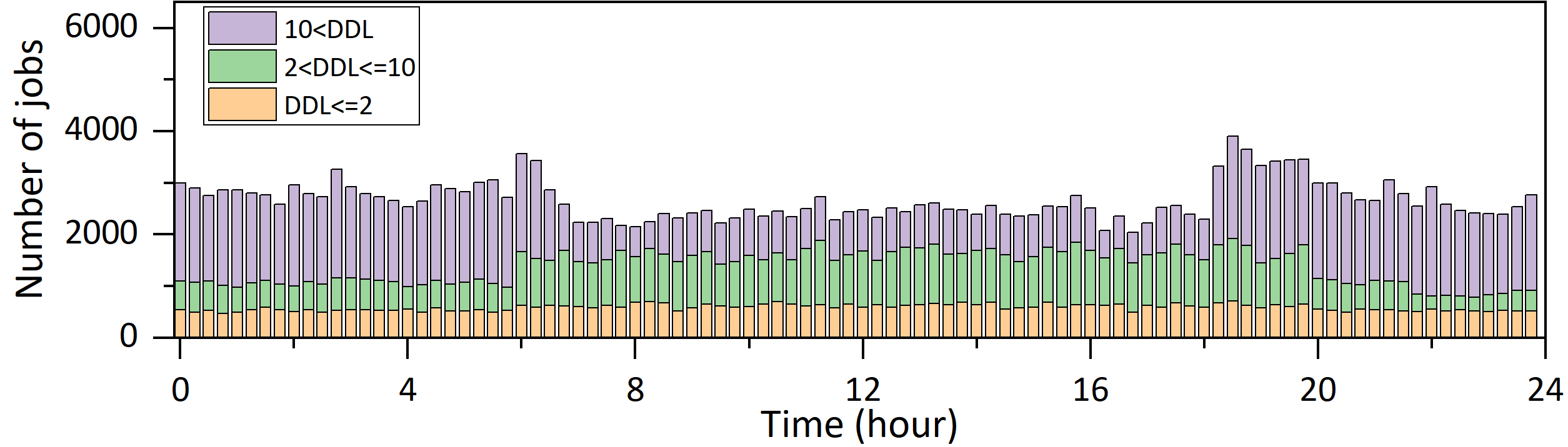}
    \caption{Highly concentrated batch job arrival patterns.}
    \label{fig:arrival_long}
\end{figure}

\begin{figure}[htbp]
    \centering
    \begin{minipage}[b]{\columnwidth}
        \includegraphics[width=\linewidth]{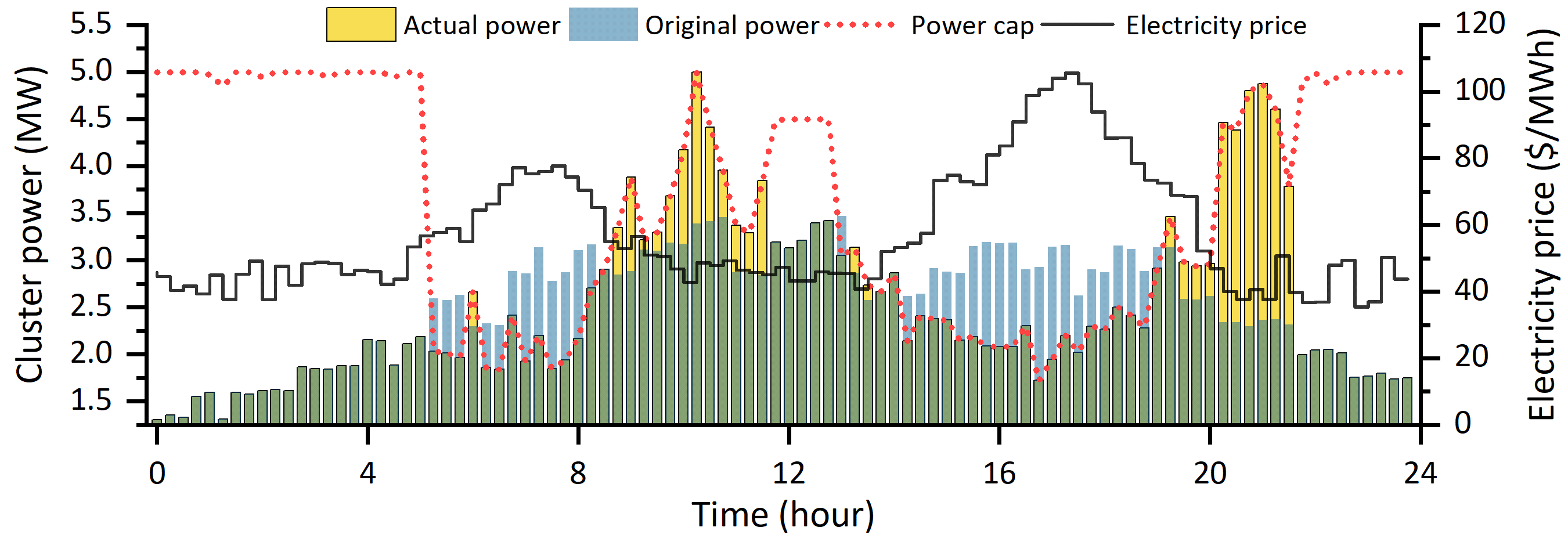}
        \label{capping_short}
    \end{minipage}
    \begin{minipage}[b]{\columnwidth}
        \includegraphics[width=\linewidth]{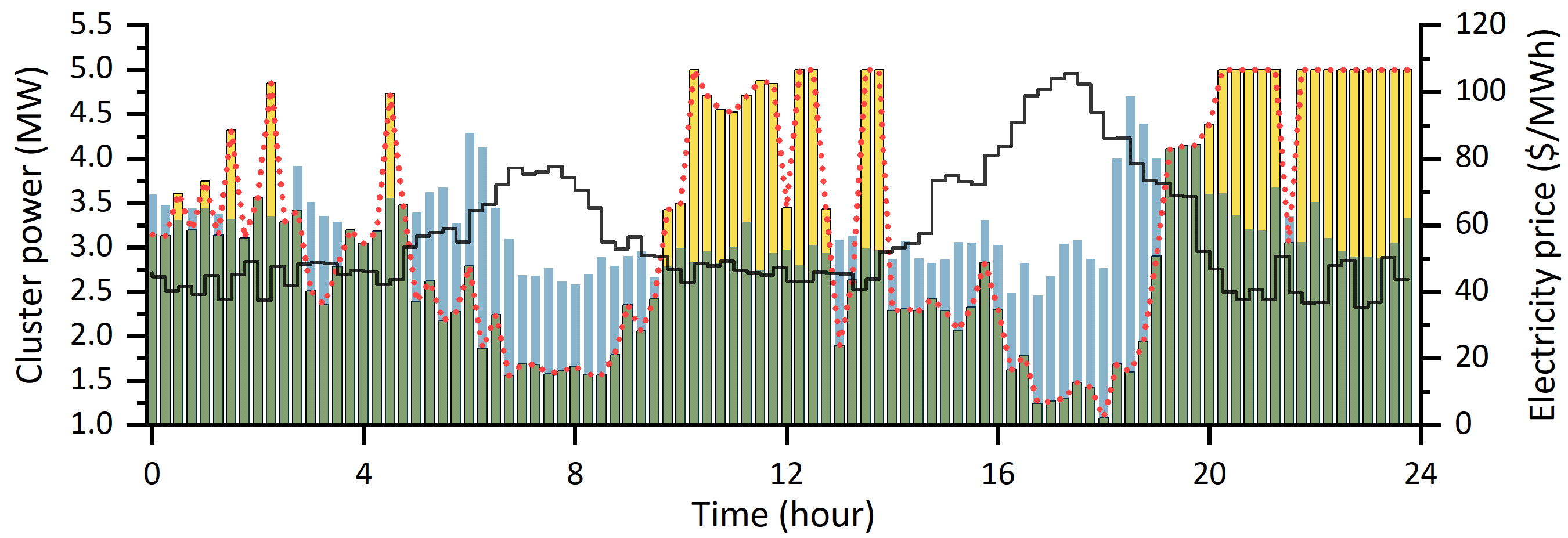}
        \label{capping_long}
    \end{minipage}
    \caption{Power capping and actual power under (a) uniformly distributed online job arrivals and (b) highly concentrated batch job arrivals.}
    \label{load patterns}
\end{figure}

We demonstrate the effectiveness of the proposed power capping method across different electricity pricing patterns, job scheduling policies and job arrival patterns in Figs. \ref{seasons}-\ref{load patterns}.

Figure \ref{seasons} illustrates the effect of the proposed power capping method on energy consumption in cloud data centers. The blue bars represent energy consumption without capping, while the yellow bars show actual consumption with power capping applied. The proposed method effectively adjusts power capping under different electricity price fluctuations. In the winter scenario as shown in Fig. \ref{seasons} (d), where electricity prices are higher and more volatile, the method tightens the power cap and defers jobs far from their DDLs during peak electricity price periods, significantly lowering power consumption and reducing energy costs. These jobs are executed during low-price periods to maintain service quality. In the summer scenario, as shown in Fig. \ref{seasons} (b), power capping is only applied during peak hours, aligning power usage with price fluctuations to minimize costs. In spring and autumn, where price fluctuations are smaller and prices are generally lower, the method applies a more relaxed power capping strategy, ensuring cost savings while maintaining service quality. Additionally, to ensure QoS for jobs with imminent deadlines, the method avoids setting excessively low power caps, maintaining a safety margin to ensure the timely execution of these jobs.

Figure \ref{different scheduler} demonstrates the compatibility of the proposed power capping method with various job scheduling algorithms. The proposed method effectively reduces energy costs in different strategies, although there is a slight increase in SLA violations. This reflects the trade-off between reducing energy costs and maintaining SLA compliance in cloud data centers. To reduce energy costs, the power capping method dynamically adjusts energy consumption based on electricity price fluctuations and job loads, often limiting power during high-price periods. However, this can impact job completion times, leading to potential SLA breaches. Despite this, the increase in SLA violations is minimal, suggesting that the method still maintains high service quality in most cases. Notably, energy cost reductions are more significant with the EDF and energy-aware policies, as these algorithms offer greater flexibility in job scheduling, enabling better alignment with power capping constraints. In contrast, the FCFS and RR algorithms lack a deadline-driven focus, raising the risk of SLA violations. As a result, the flexibility for energy adjustments is more limited to ensure SLA compliance in these cases.

To enhance generality, we further evaluate two representative job arrival patterns.
\begin{itemize}
    \item \textbf{Uniformly distributed online jobs.} This pattern, common in globally distributed SaaS platforms, features latency-sensitive and short-lived jobs arriving more frequently during daytime hours~\cite{iqbal2010sla}. The corresponding arrival profile is shown in Fig.~\ref{fig:arrival_short}.

    \item \textbf{Highly concentrated batch jobs.} This pattern involves large volumes of delay-tolerant jobs arriving periodically~\cite{ding2018emission,cheng2018analyzing}, typical in enterprise private clouds and governmental data centers. The arrival profile is illustrated in Fig.~\ref{fig:arrival_long}.
\end{itemize}

As shown in Fig. \ref{load patterns}, the proposed method effectively adjusts the power load profile across all scenarios by deferring job execution away from peak-price periods, thereby achieving significant cost savings. In the case dominated by short-deadline jobs, strict latency constraints reduce scheduling flexibility, as lowering the power cap would incur prohibitive delay penalties. In contrast, long-deadline jobs provide greater temporal flexibility, enabling more substantial reshaping of the power profile. These jobs are shifted to low-price periods, allowing effective peak shaving during nearly all high-price intervals. Overall, the results confirm that the proposed framework performs effectively under different types of job pattern and offers a greater optimization space when jobs exhibit higher tolerance to execution delay.

\vspace{-3mm}
\subsection{Fast Learning for the Adaptive Power Capping Policy}
This subsection conducts several numerical experiments to demonstrate the fast learning performance and adaptability of the proposed method by comparing it with commonly used model-free RL algorithms, PPO, SAC, DQN and IQN. 

\begin{figure}[h]
    \centering
    \includegraphics[width=8cm]{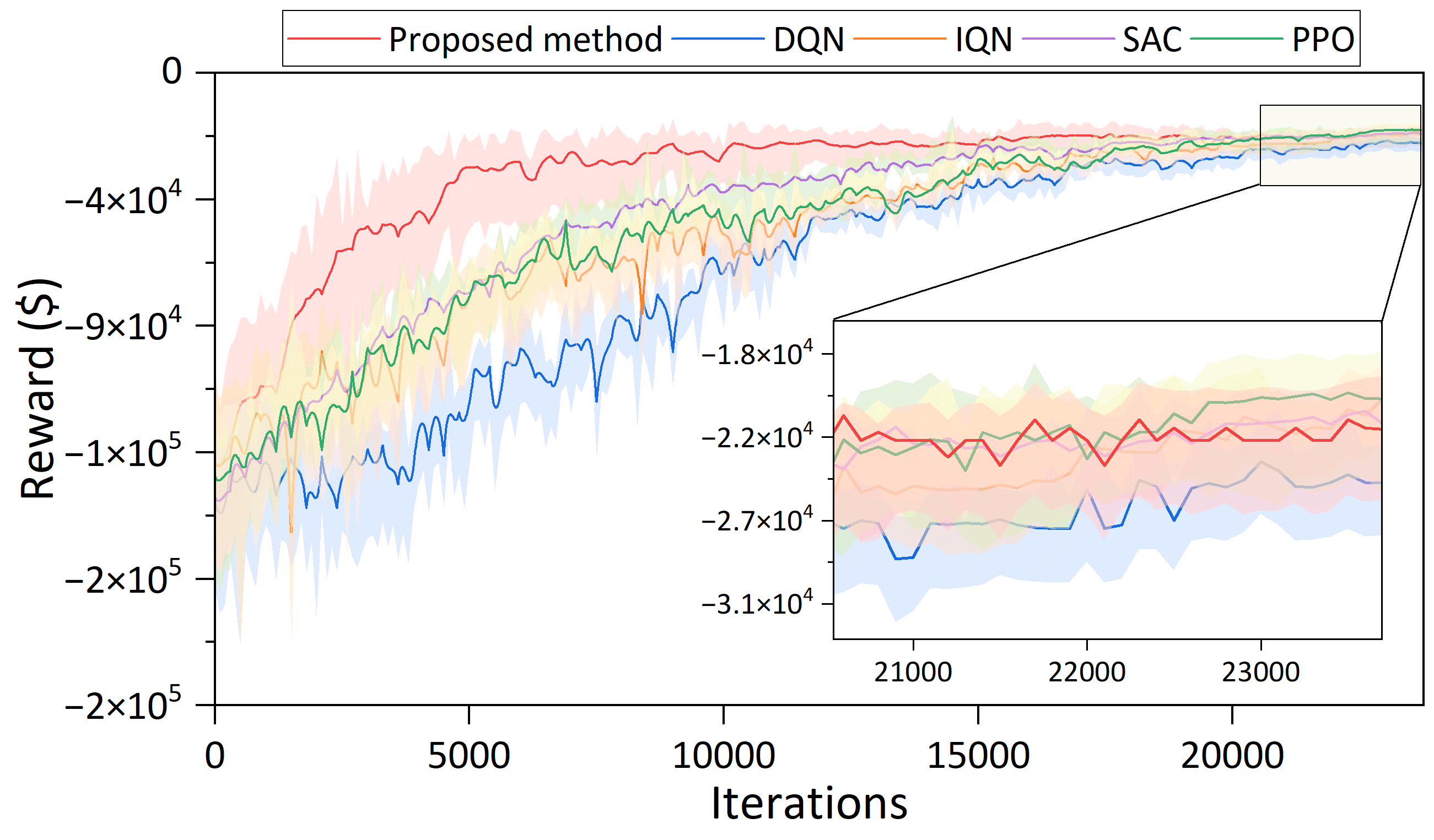}
    \caption{Training rewards of the proposed method and benchmark models.}
    \label{fastlearning}
\end{figure}

We present the training results of these methods, as illustrated in Fig. \ref{fastlearning}. The figure  shows that the proposed method exhibits a rapid and noticeable improvement compared to the other methods. After approximately 2,500 iterations, the proposed method surpasses the others in reward performance. This is because the proposed method can learn an environment transition model based on data obtained from real-world interactions. The model then generates virtual trajectories to optimize decision-making. This process reduces the need for direct interactions with the environment, enabling the method to fast learn more effective policies with fewer interactions. Furthermore, at convergence,  the performance of the proposed method closely matches that of PPO, with a total reward difference of only 6.47\%. However, it reaches this performance 2.17 times faster, highlighting its advantages in learning speed and sample efficiency.
\vspace{-5mm}
\begin{figure}[htbp]
    \centering
    \begin{minipage}[b]{\columnwidth}
        \includegraphics[width=\linewidth]{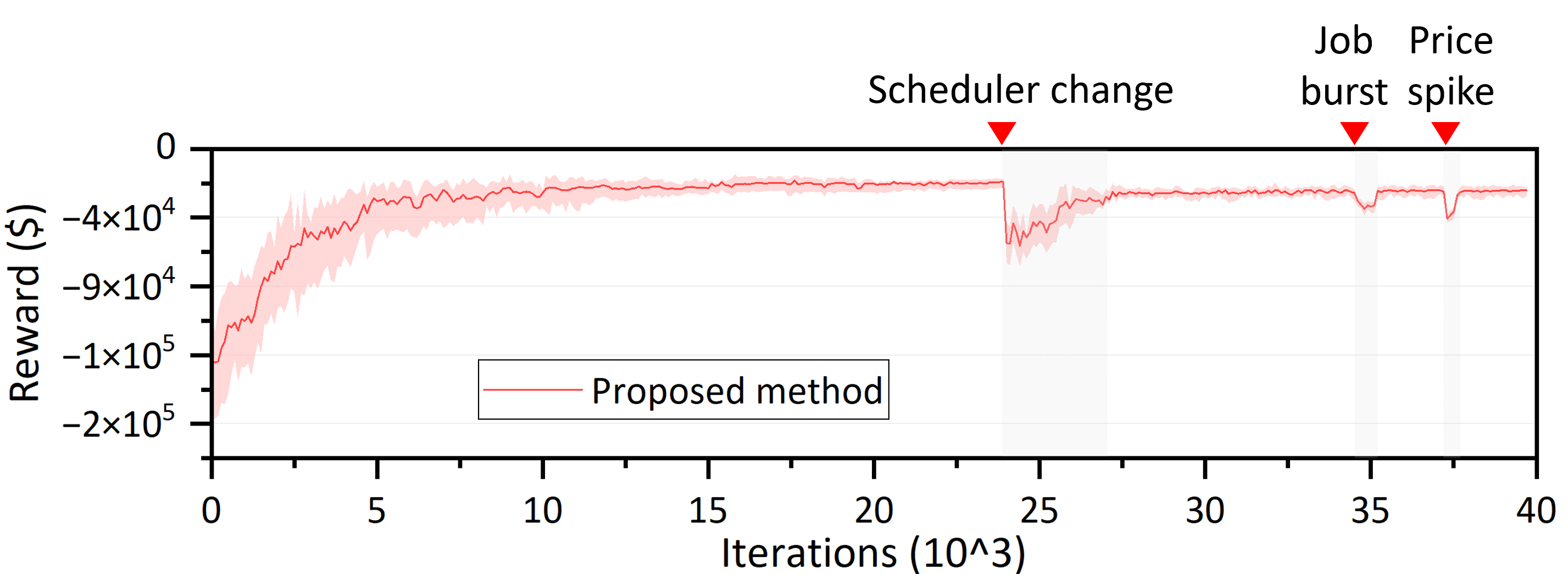}
        \label{spring}
    \end{minipage}
    \begin{minipage}[b]{\columnwidth}
        \includegraphics[width=\linewidth]{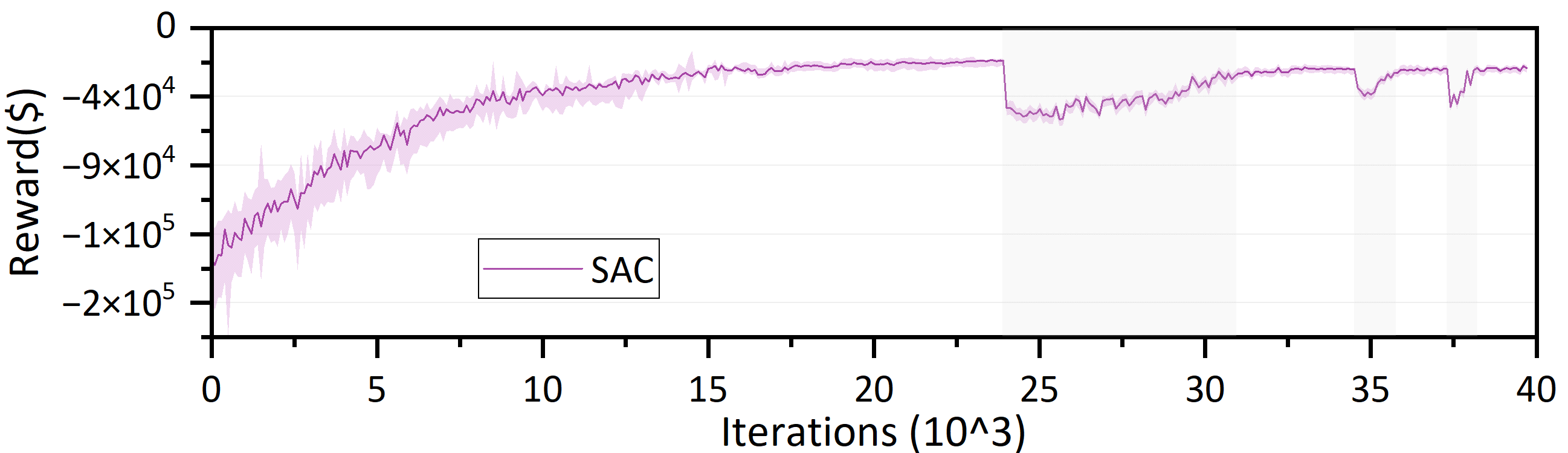}
        \label{summer}
    \end{minipage}
    \begin{minipage}[b]{\columnwidth}
        \includegraphics[width=\linewidth]{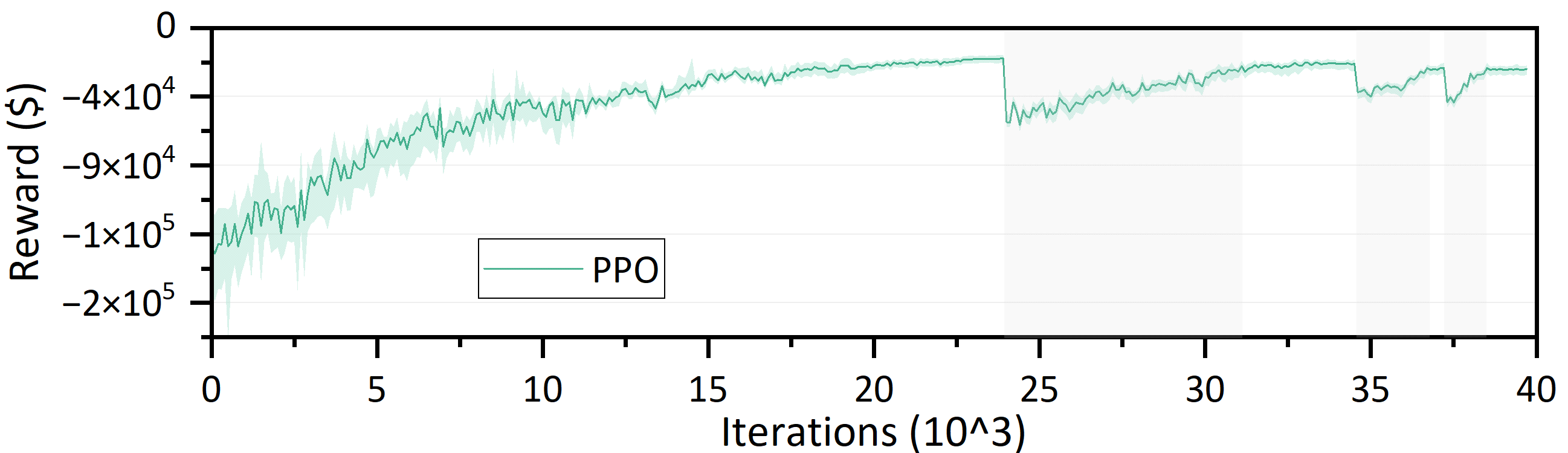}
        \label{autumn}
    \end{minipage}
    \begin{minipage}[b]{\columnwidth}
        \includegraphics[width=\linewidth]{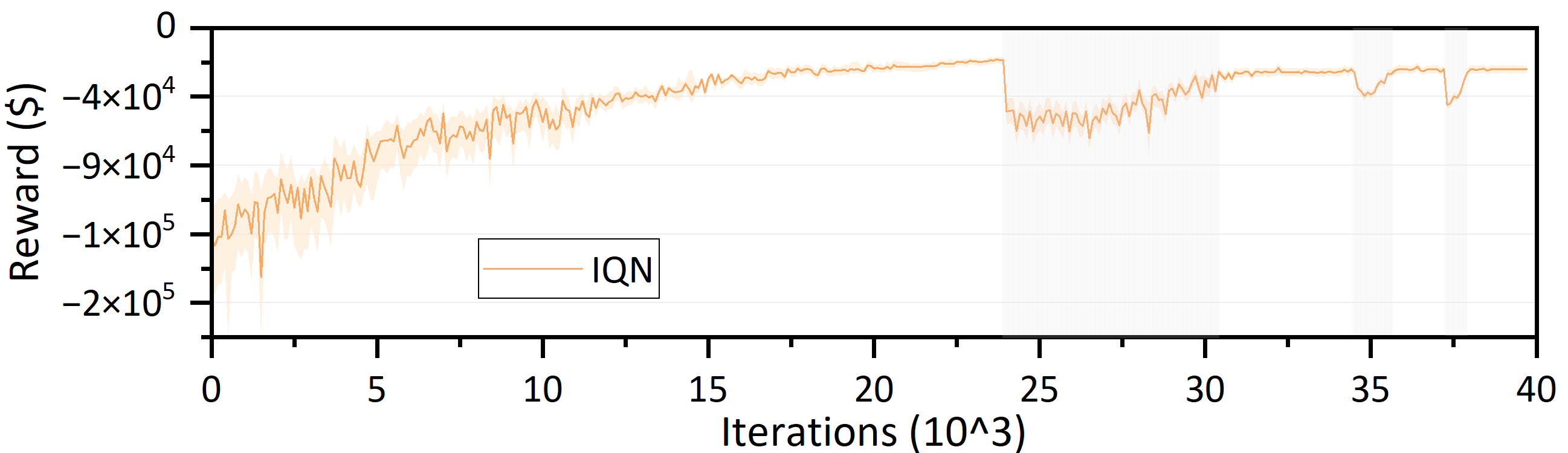}
        \label{winter}
    \end{minipage}
    \begin{minipage}[b]{\columnwidth}
        \includegraphics[width=\linewidth]{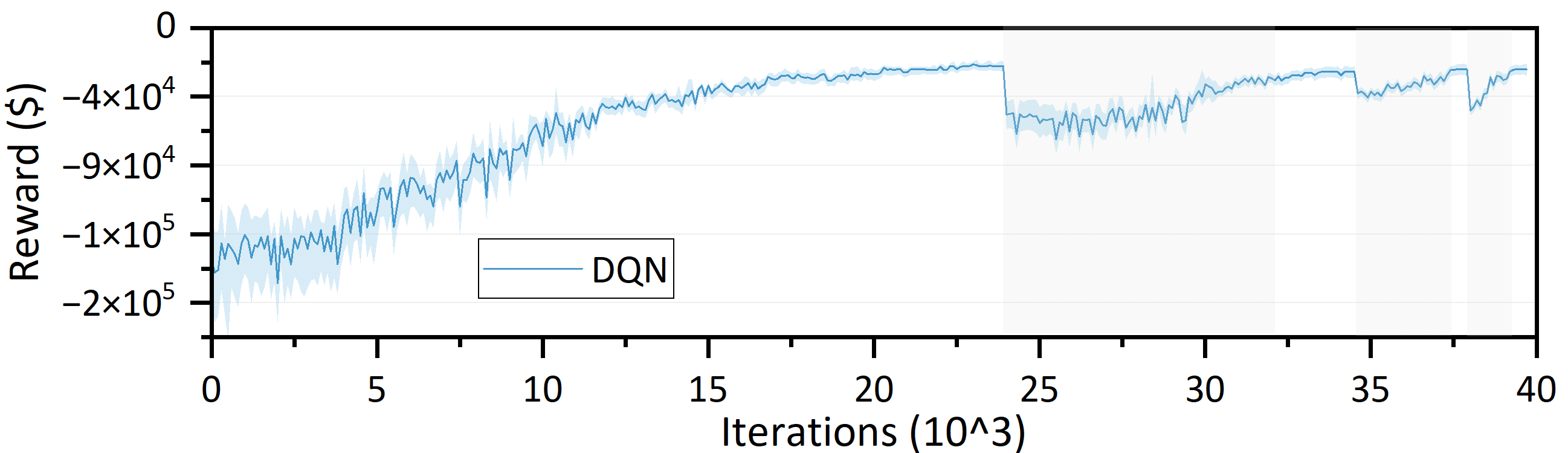}
        \label{winter}
    \end{minipage}
    \caption{Training performance under sudden environmental changes.}
    \label{fast recorvery}
    \vspace{-3mm}
\end{figure}

Figure \ref{fast recorvery} shows the fast learning and adaptive feature of the proposed power capping method in a dynamic computing environment. Initially, the power capping policy is trained through interactions with an EDF policy. However, at a certain point, the type of job arriving at the data center changes abruptly, accompanied by different SLA requirements, resulting in a shift in scheduling policy from EDF to FCFS. Subsequently, a job burst occurs, followed by a price spike. As indicated in Fig. \ref{fast recorvery}, there is a noticeable drop in reward as the model encounters a new environmental changes. Despite this sudden disruption, the proposed method quickly adapts, learning the FCFS scheduling pattern and load variation dynamics within a short time, ultimately achieving new stable performance. In contrast, the other methods exhibit longer adaptation periods, with rewards that remain suboptimal for a long duration. This lag stems from their reliance on numerous interactions with the environment to relearn effective strategies, resulting in extended periods of reduced performance. Furthermore, during the job burst, the proposed method experiences a smaller drop in reward compared to the other methods. This is attributed to the ability of the environment model to partially simulate changes in the environment, thereby mitigating the negative impacts of such disruptions. The adaptability of the proposed method highlights its capability for online decision-making and adjustment. This is crucial for cloud data centers, where scheduling algorithms may change based on operational needs or job demands, providing practical value in real-world applications.

\subsection{Power Capping under Uncertainty in Cloud Data Centers}
In this subsection, we evaluate the performance of the proposed method in handling uncertainty by comparing it with various benchmarks, including learning-based, rule-based and forecast-based methods.

\begin{figure}[h]
    \centering
    \includegraphics[width=8.5cm]{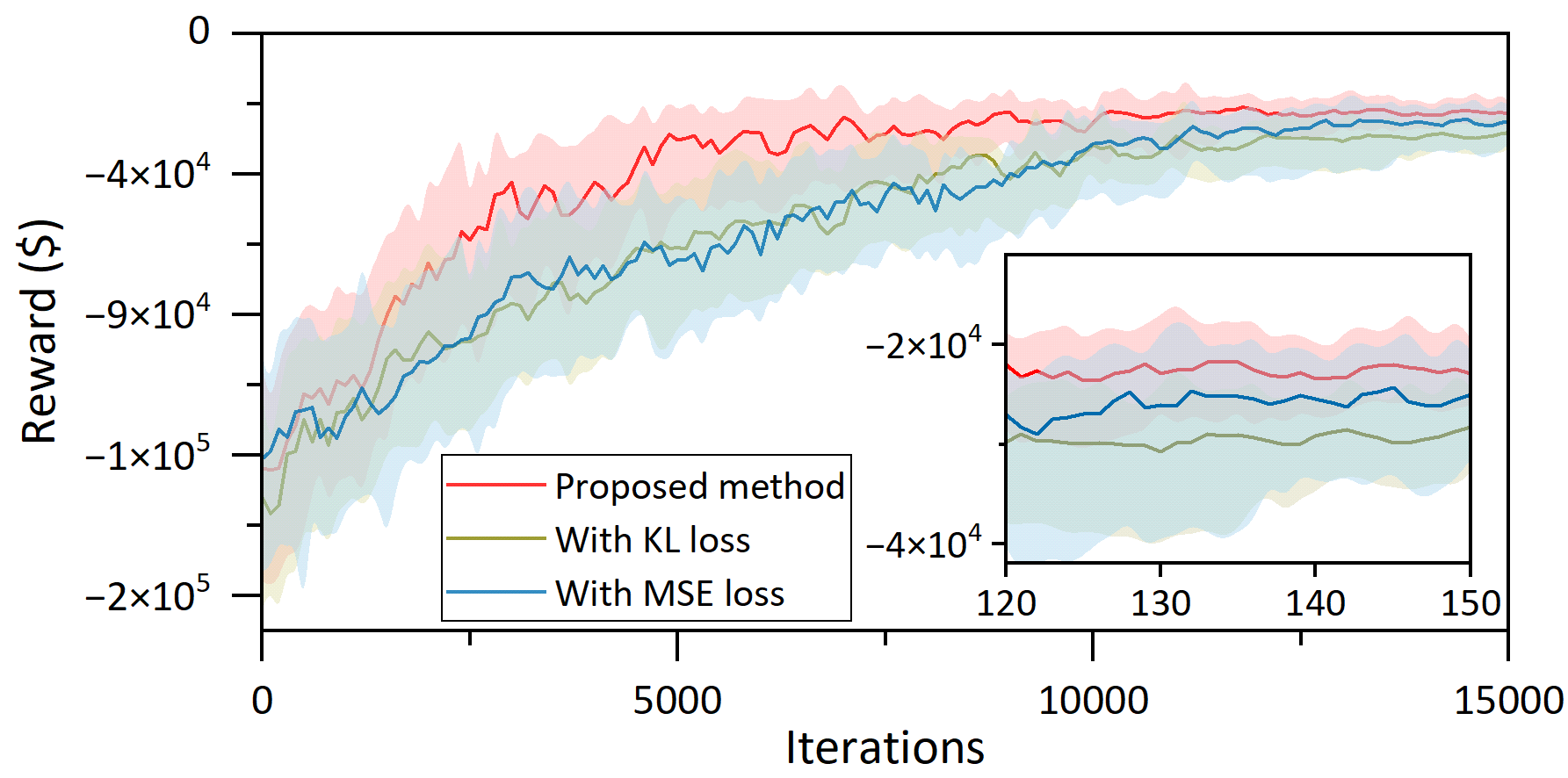}
    \caption{Training rewards of MBRL methods with different uncertainty considerations.}
    \label{uncertain RL}
    \vspace{-3mm}
\end{figure}

We perform a comparative analysis with other MBRL methods to assess the impact of integrating uncertainty and decision-making information during the transition dynamics learning stage on performance. The KL divergence loss method accounts for uncertainty but excludes decision-making information \cite{hafner2019learning}, while the MSE loss method incorporates decision-making knowledge but simplifies the handling of uncertainty \cite{farahmand2016value}. As illustrated in Fig. \ref{uncertain RL}, the proposed method exhibits a faster convergence rate and achieves a higher final reward compared to the alternative approaches. Unlike the KL loss method, the proposed method incorporates decision-making context to enhance its capacity to learn dynamic transitions. The method with KL loss struggles to fully capture the influence of dynamic environmental changes on decision-making because of the lack of decision-making information, resulting in slower convergence and lower reward levels. Similarly, while the method with MSE loss converges slightly faster than the KL loss method, it exhibits greater variance. This instability suggests that although decision-making information supports environmental exploration, the method's simplistic treatment of uncertainty leaves it more fragile to environmental changes. By integrating both uncertainty and decision-making knowledge within a unified optimization framework, the proposed method demonstrates a superior ability to handle environmental uncertainties, yielding enhanced performance outcomes.

We further demonstrate the effectiveness of the proposed power capping method in addressing uncertainty in cloud data centers by comparing it with rule-based and forecast-based benchmarks.

\textbf{Rule-based method:} This method determines the power cap using predefined rules that consider job types, current load utilization, and electricity prices \cite{wu2013classified}.

\textbf{Forecast-and-optimize method:} This method forecasts next-day job arrivals, followed by a one-shot optimization of the power cap based on the forecast to minimize the carbon emission cost. For comparison in this paper, we substitute the carbon emission factor data with electricity price data \cite{radovanovic2022carbon}.

\begin{figure}[h]
    \centering
    \includegraphics[width=9cm]{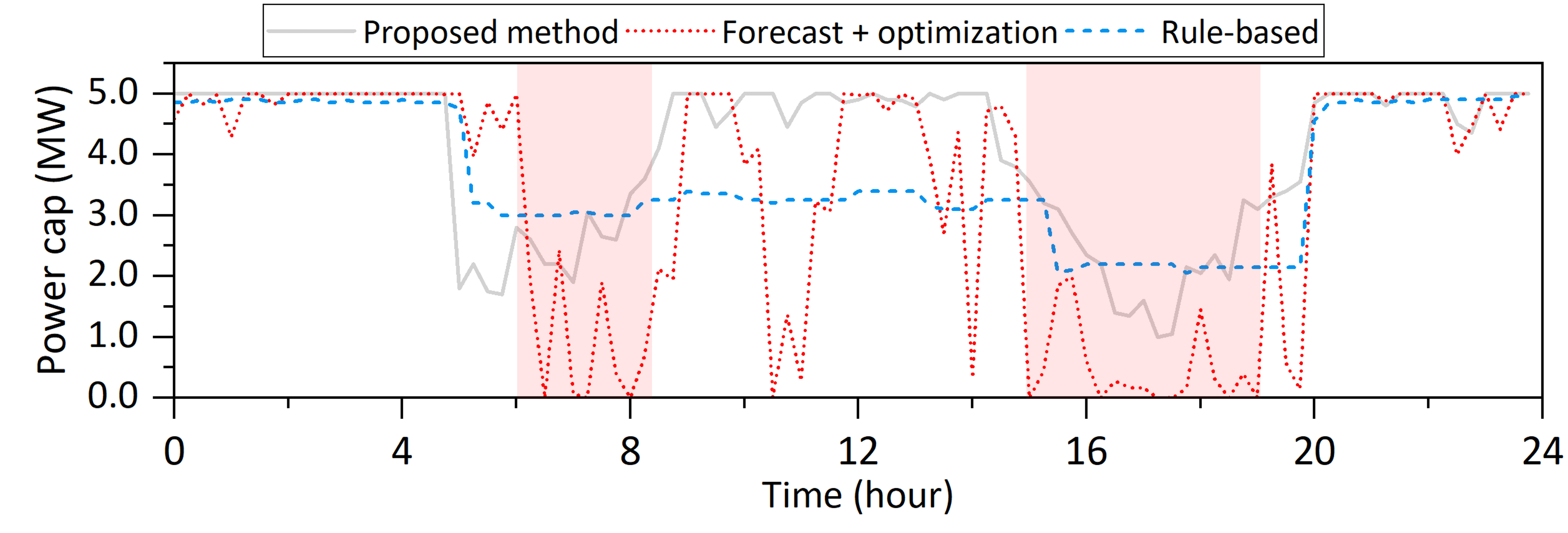}
    \vspace{-6mm}
    \caption{Comparison of power capping methods under uncertain environments.}
    \label{uncertain capping}
\end{figure}

Figure \ref{uncertain capping} shows the power capping results of the three methods in uncertain environments. The rule-based method (blue dashed line) lacks flexibility in responding to electricity price fluctuations and job arrivals, often failing to effectively adjust to changes. Its overall power capping curve is relatively flat, demonstrating insufficient responsiveness. The forecast-and-optimize method better tracks price trends. However, it lacks consideration for uncertainties and dynamically changing environments, often resulting in excessively low power caps (as shown in the red-shaded areas). These restrictive caps hinder job execution, leading to increased delays and a higher risk of exceeding SLA requirements. In contrast, the proposed method adapts to price variations while considering uncertainties in job arrivals and execution, thus avoiding overly restrictive caps that could compromise job completion.

\begin{figure}[h]
    \centering
    \includegraphics[width=9cm]{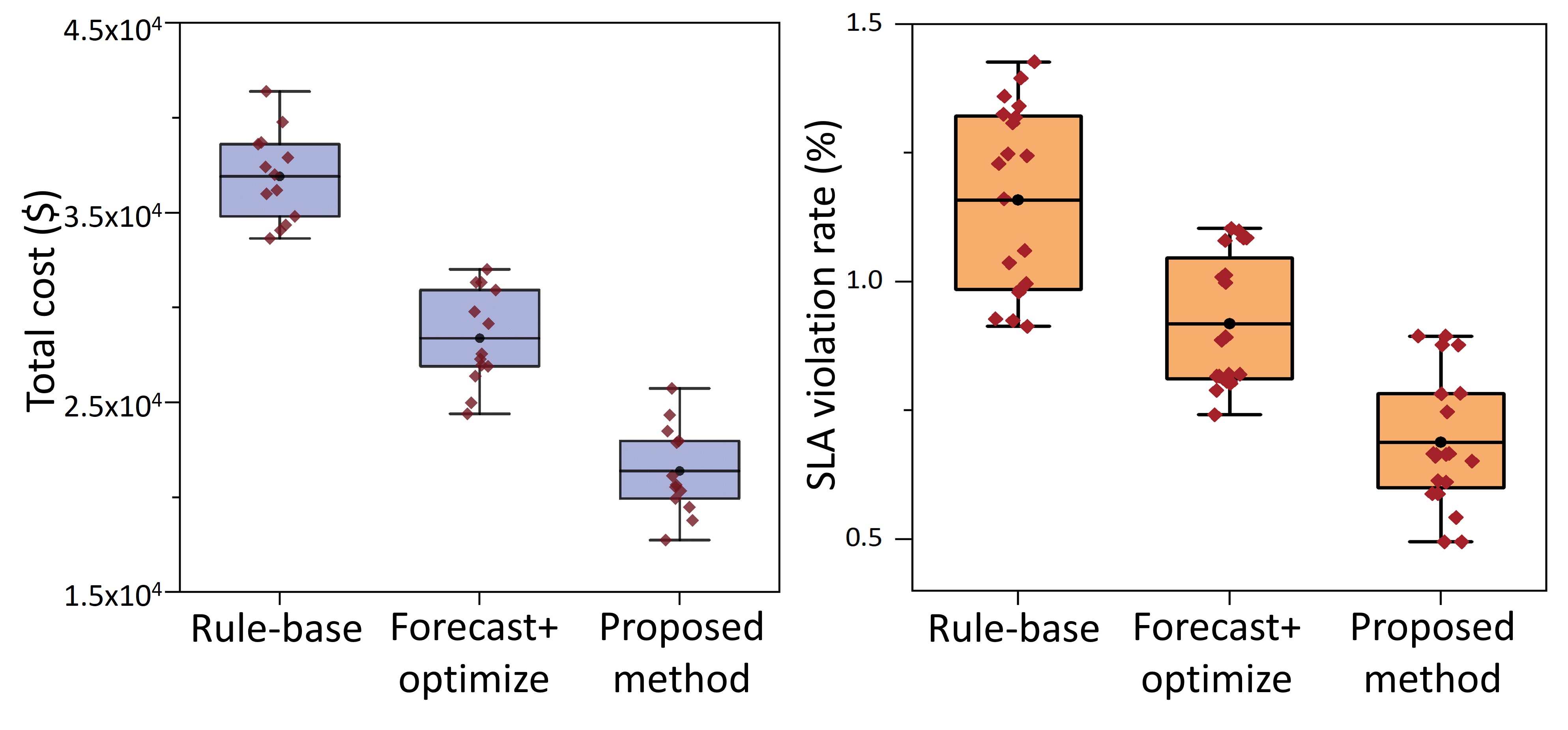}
    \caption{Total cost and SLA violation rate across different power capping methods.}
    \label{cost sla}
    \vspace{-3mm}
\end{figure}

Figure \ref{cost sla} illustrates the distribution of total cost and SLA violation rates for the three methods in the uncertain cloud data center environment. The rule-based method, which relies solely on predefined rules, has limited adaptability to dynamic fluctuations in jobs and electricity prices, resulting in higher energy consumption and SLA violations. Compared to the rule-based method, the forecast-and-optimize method offers moderate improvements, but it highly depends on the accuracy of the forecast. Furthermore, since this method performs a day-ahead one-shot optimization, it ignores job scheduling policy differences, limiting the flexibility to adapt to changes and resulting in higher energy consumption and SLA violations. In contrast, the proposed method dynamically learns and adapts to environmental uncertainties, allowing adaptive power cap adjustments. This adaptability reduces energy costs and SLA violation rates, demonstrating superior performance under uncertain conditions.

\subsection{Sensitivity Analysis}
In this subsection, we analyze the performance of our proposed method in four common SLA configurations with varying penalty mechanisms to assess their impact \cite{SLA1,overby2017sla}.

\textbf{Case I:} Default setting. Any job missing its deadline is immediately penalized with a fixed cost per time step until completion.

\textbf{Case II:} A more stringent SLA where the penalty increases linearly with the duration of the delay.

\textbf{Case III:} A relaxed SLA that allows for a short grace period (e.g., two steps) during which delays do not incur penalties. Only delays beyond this window are penalized.

\textbf{Case IV:} Same as in Case I, but the penalty per time step is reduced by half.

\begin{figure}[h]
    \centering
    \includegraphics[width=\linewidth]{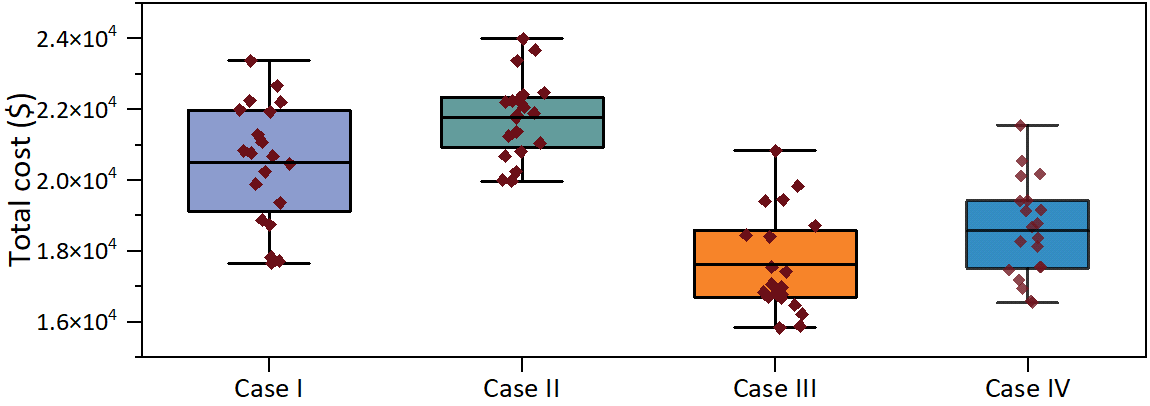}
    \caption{Total cost across different SLA configurations.}
    \vspace{-6mm}
\end{figure}
\begin{figure}[h]
    \centering
    \includegraphics[width=\linewidth]{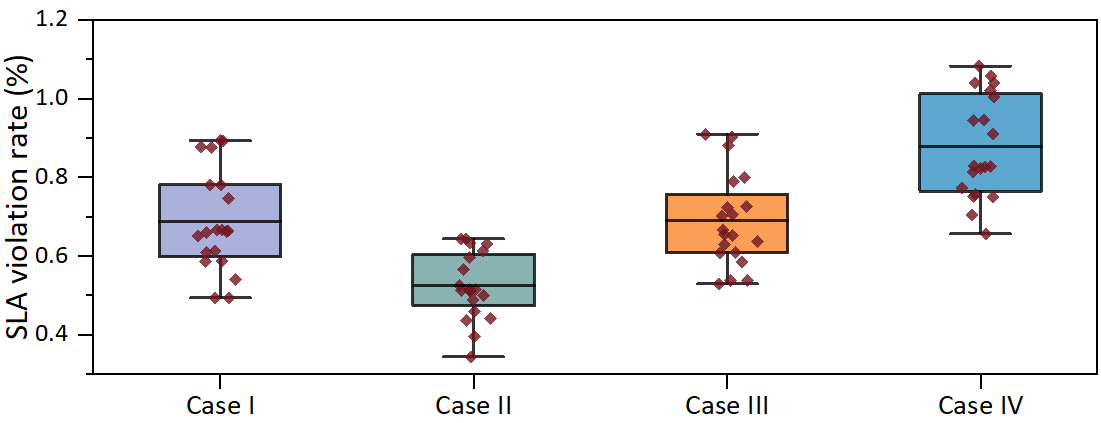}
    \caption{Violation rate across different SLA configurations.}
    \vspace{-3mm}
\end{figure}

We compare and analyze the total costs and violation rates in different SLA configurations. Case II, due to the cumulative effect of penalties, imposes significant costs for missed deadlines, resulting in the lowest violation rate. However, substantial penalties for violations lead to higher overall costs. In contrast, Case III, by introducing a grace period during which delays incur no penalties, offers greater flexibility in job scheduling, thereby reducing total costs, although the violation rate increases slightly. On the other hand, Case IV applies lighter penalties for violations, permitting more delays. Consequently, while total costs decrease marginally, the number of violations increases significantly. In general, the proposed approach strikes a balance between cost-effectiveness and violation rates.

\section{Conclusion}
This paper proposes an adaptive power capping approach for cloud data centers aimed at optimizing energy consumption in uncertain and dynamic environments while achieving fast strategy learning with fewer interactions to effectively address the costly learning process. We first formulate the adaptive power capping framework as a POMDP model to capture uncertainties in job arrivals, electricity prices, and ever-changing scheduling policies. To address these uncertainties, we introduce an uncertainty-aware MBRL method that incorporates both environmental and decision uncertainties into the model learning stage. Moreover, we derive the optimality gap of the power capping policy with finite iterations to ensure an acceptable optimization performance. Finally, we conduct numerical experiments to demonstrate the effectiveness of the proposed method in managing energy costs and maintaining SLA compliance across various scheduling policies and uncertain conditions, offering a practical solution for cloud data centers. It is worth noting that future extensions may incorporate energy storage systems (ESS) to further enhance the effectiveness of power capping strategies. Beyond the reduction of electricity costs for data centers, ESS can enhance operational resilience by mitigating sudden power ramps, supporting grid stability, and facilitating the integration of renewable energy.

\appendices

\section{Upper Bound of Model Learning Error Term}
We consider the term \(\gamma_K \sum\limits_{k=0}^{K-1} \alpha_k z_0 A_k \Delta_k\) in (24), which is related to the environment model learning error \(\Delta_k\). Using the Neumann series expansion for \(\mathcal{I}_K^{-1}\) and the definition of \(\overline{c}_{z_0, z_k}(k)\) in Definition 1, we derive:
\begin{align}
z_0 A_k &= \frac{1}{2} z_0 \sum_{m \geq 0} \gamma^m (\mathcal{P}^{\pi_K})^m 
\left[ (\mathcal{P}^{\pi^*})^{K-k} + \mathcal{P}^{\pi_K}(\mathcal{P}^{\pi^*})^{K-k-1} \right] \nonumber \\
&\leq \sum_{m \geq 0} \gamma^m \bar{c}_{z_0, z_K}(K-k+m).
\end{align}

Expanding \(\alpha_k\) and \(\gamma_K\), combining them with (28), and applying variable substitution and Definition 1, we obtain:
\begin{align}
\gamma_K \sum_{k=0}^{K-1} \alpha_k z_0 A_k &\leq 2 \gamma \sum_{k=0}^{K-1} \sum_{m \geq 0} \gamma^{K-k-1+m} \bar{c}_{z_0,z_K}(K - k + m) \nonumber \\
&= 2 \gamma \sum_{m \geq 0} \sum_{k'=1}^{K-1} \gamma^{k'-1+m} \bar{c}_{z_0,z_K}(k' + m) \nonumber \\
&= 2 \gamma \sum_{n \geq 1} n \gamma^{n-1} \bar{c}_{z_0,z_K}(n) \\
&= 2 \gamma \frac{\bar{C}(z_0,z_K)}{(1 - \gamma)^2}. \nonumber
\end{align}

To simplify the expression, we take the maximum of \(\Delta_k\) over \(k = 0, \ldots, K - 1\), and move it outside the summation:
\begin{equation}
\gamma_K \sum_{k=0}^{K-1} \alpha_k z_0 A_k \Delta_k \leq 2 \gamma \frac{\overline{C}(z_0, z_K)}{(1 - \gamma)^2} \max_{0 \leq k \leq K-1} \Delta_k.
\end{equation}

Thus, we derive an upper bound on the model learning error term \(\Delta_k\), demonstrating that the final policy error is partially influenced by the precision of the model learning process. This result suggests that controlling the model error within a small range contributes to improving the reliability of the final policy.

\section{Upper Bound of the Initial Distribution Distance Term}
Next, we analyze the term in (24) related to the distance between the initial state-action value distribution \(z_0\) and the optimal state-action value distribution \(Z^*\), expressed as \(\gamma_K \alpha_K z_0 A_K \mathcal{W}(Z^*, z_0)\).

For the proposed POMDP, the rewards are finite due to the inherent characteristics of the power capping problem, meaning that single-step rewards are bounded by \(R_{\text{max}}\). From the definitions of state-action and state value functions, we have \(Q(s_t, a_t) \leq V(s_t)_{\text{max}}\) for any time step and state. For distributional state-action value functions, the distance between any corresponding quantiles of two distributions is bounded by \(2V_{\text{max}}\) \cite{farahmand2018iterative}.

Given that \(z_0\) and \(A_K\) lie within \((0, 1)\), it follows that:
\begin{align}
z_0 A_K \mathcal{W}(Z^*, z_0) &\leq 2V_{\text{max}}.
\end{align}
Using the relationship \(V_{\text{max}} = \sum\limits_{t=0}^\infty \gamma^t R_{\text{max}} = \frac{R_{\text{max}}}{1 - \gamma}\), expanding and substituting \(\gamma_K\) and \(\alpha_K\) into (31), we derive:
\begin{equation}
\gamma_K \alpha_K z_0 A_K \mathcal{W}(Z^*, z_0) \leq 4 \gamma^{K+1} R_{\text{max}}.
\end{equation}
From (32), we observe that the policy error bound is partially influenced by the maximum reward bound. Since \(0 < \gamma < 1\), \(\gamma^{K+1}\) decreases exponentially, implying that as \(K\) increases, the policy converges toward optimality, aligning with intuitive understanding.

\bibliographystyle{IEEEtran} 

\bibliography{ref} 

\begin{IEEEbiography}[{\includegraphics[width=1in,height=1.25in,clip,keepaspectratio]{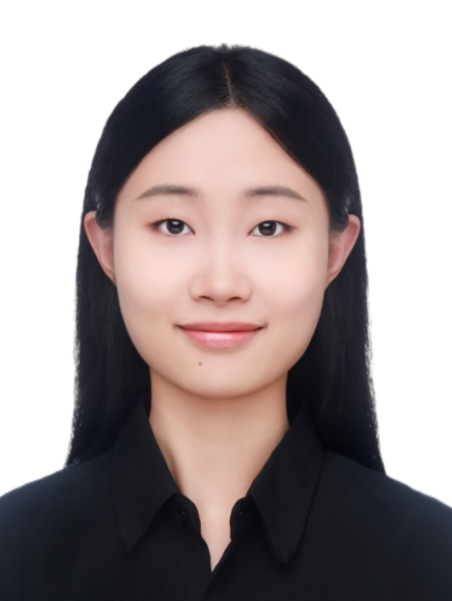}}]{Yimeng Sun}
(Graduate Student Member, IEEE) received the B.S. degree in the electrical engineering and automation in 2021 from North China Electric Power University, China, where she is currently pursuing the Ph.D. degree in electrical engineering with School of Electrical and Electronic engineering. Her current research interests include power system optimization and demand side management. 
\end{IEEEbiography}

\begin{IEEEbiography}[{\includegraphics[width=1in,height=1.25in,clip,keepaspectratio]{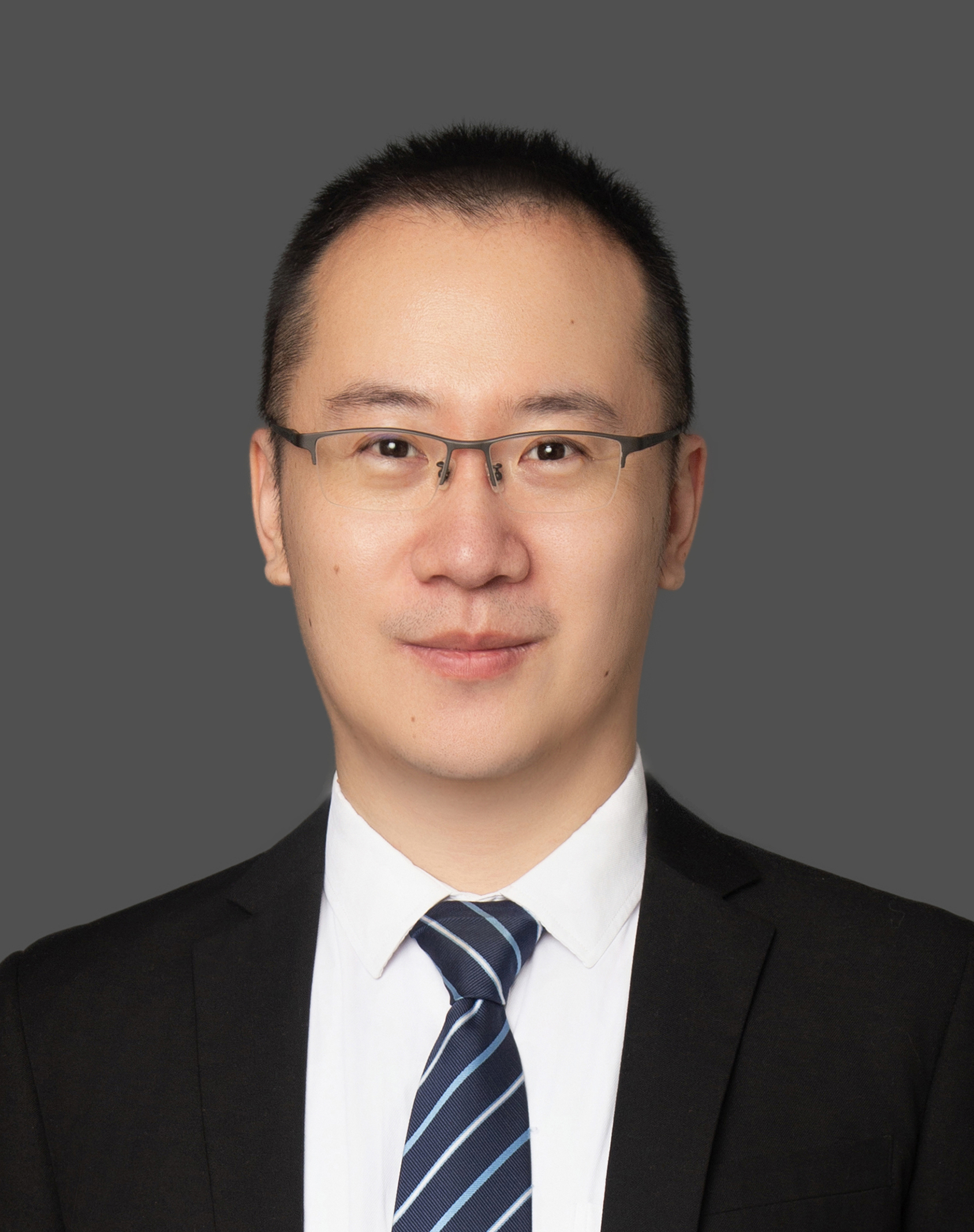}}]{Zhaohao Ding}
(Senior Member, IEEE) received the B.S. degree in electrical engineering and the B.A. degree in finance from Shandong University, Jinan, China, in 2010, and the Ph.D. degree in electrical engineering from the University of Texas at Arlington, Arlington, TX, USA, in 2015. He is currently a Professor with North China Electric Power University, Beijing, China. His research interests include power system planning and operation, power market, distributed resource management, and electric transportation system.
\end{IEEEbiography}

\begin{IEEEbiography}[{\includegraphics[width=1in,height=1.25in,clip,keepaspectratio]{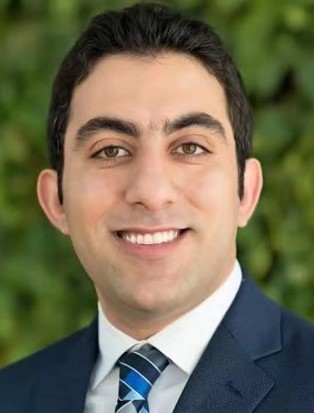}}]{Payman Dehghanian} (Senior Member, IEEE) received the B.Sc. degree in electrical engineering from the University of Tehran, Tehran, Iran, in 2009, the M.Sc. degree in electrical engineering from the Sharif University of Technology, Tehran, in 2011, and the Ph.D. degree in electrical engineering from Texas A\&M University, College Station, TX, USA, in 2017. In 2018, he joined the Department of Electrical and Computer Engineering, The George Washington University, Washington, DC, USA, where he is currently an Associate Professor. His research interests include power systems reliability and resilience assessment, data-informed decision making in power and energy systems, and smart electricity grid applications. He was the recipient of the 2014 and 2015 IEEE Region 5 Outstanding Professional Achievement Awards, 2015 IEEE-HKN Outstanding Young Professional Award, 2021 Early Career Award from the Washington Academy of Sciences, 2022 George Washington University’s Early Career Researcher Award, 2022 IEEE IAS Electrical Safety Committee’s Young Professional Achievement Award, and 2022 IEEE IAS Outstanding Young Member Service Award. In 2015 and 2016, he was selected among the World’s Top 20 Young Scholars for Next Generation of Researchers in electric power systems.
\end{IEEEbiography}

\begin{IEEEbiography}
[{\includegraphics[width=1in,height=1.25in,clip,keepaspectratio]{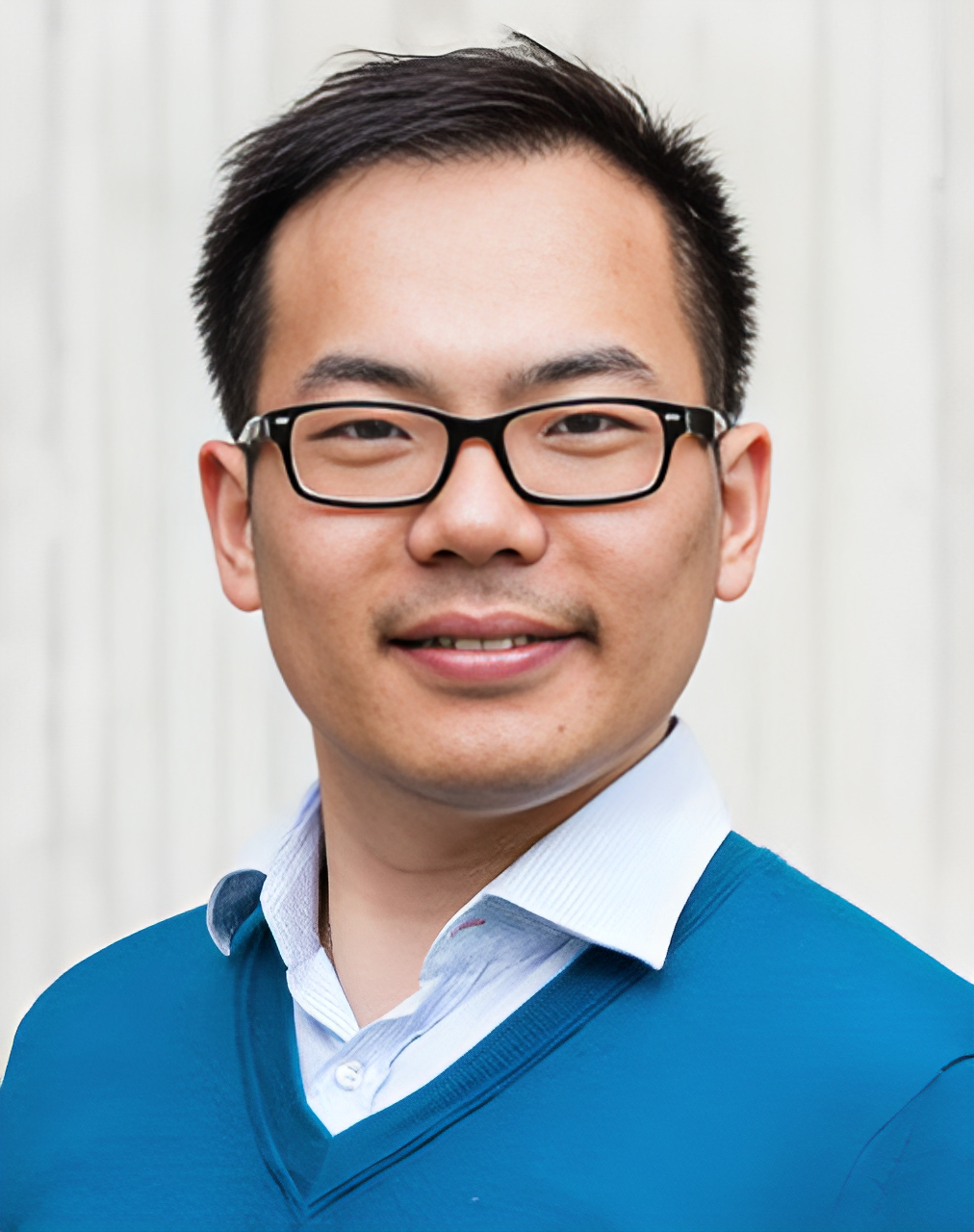}}]{Fei Teng}
(Senior Member, IEEE) received the B.Eng. degree in electrical engineering from Beihang University, China, in 2009, and the M.Sc. and Ph.D. degrees in electrical engineering from Imperial College London, U.K., in 2010 and 2015, respectively. He is currently a Senior Lecturer with the Department of Electrical and Electronic Engineering. His research focuses on the power system operation with high penetration of Inverter-Based Resources (lBRs) and the Cyber-resilient and Privacy-preserving cyber-physical power grid.
\end{IEEEbiography}

\end{document}